\documentclass[twocolumn,aps,pra,superscriptaddress]{revtex4}

\usepackage{graphicx}
\usepackage{amsmath,amssymb}
\usepackage{multirow}
\usepackage{color}
\usepackage[colorlinks=true,citecolor=blue,urlcolor=blue]{hyperref}
\usepackage{bbold}
\usepackage{pdflscape}
\usepackage{afterpage} 
\usepackage{capt-of}
\usepackage{xcolor}
\usepackage{hyperref}
\usepackage{setspace}
\usepackage{footnote}
\usepackage{braket}
\usepackage{appendix}
\usepackage{lipsum}
\usepackage[colorinlistoftodos]{todonotes}
\usepackage{blindtext}
\usepackage{etoolbox}
\newcommand{\nn}{\texttt{n}} %refraction index

\begin{document}
\title{Exploring entanglement in open cavity parametric oscillators: from triply to doubly resonant cavities}
%%%%%%%%%%%%%%%%%%%%%%%%%%%%%%%%%%%%%%%%%%%%%%%%%%%%%%%%%%%%%%%%%%%
%\footnote{Both authors contributed equally for the present paper\label{footnote}}
\author{B\'arbara Abigail Ferreira Ribeiro}
\thanks{B.A.F.R. and R.B.A. contributed equally for this project.}
\affiliation{Instituto de F\'isica, Universidade de S\~ao Paulo, P O Box 66318, 05315-970 S\~ao Paulo, Brazil}
\author{Rayssa Bruzaca de Andrade}
\email{rabda@fysik.dtu.dk}
\affiliation{Instituto de F\'isica, Universidade de S\~ao Paulo, P O Box 66318, 05315-970 S\~ao Paulo, Brazil}
%\author{Paulo Nussenzveig}
%\affiliation{Instituto de F\'isica, Universidade de S\~ao Paulo, P O Box 66318, 05315-970 S\~ao Paulo, Brazil}
\author{Marcelo Martinelli}
\affiliation{Instituto de F\'isica, Universidade de S\~ao Paulo, P O Box 66318, 05315-970 S\~ao Paulo, Brazil}
\author{Breno Marques}
\email{breno.marques@ufabc.edu.br}
\affiliation{Instituto de F\'isica, Universidade de S\~ao Paulo, P O Box 66318, 05315-970 S\~ao Paulo, Brazil}
\affiliation{Ci\^encias  Naturais  e  Humanas,  Universidade  Federal  do  ABC  -  UFABC,  Santo  Andr\'e,  Brazil}

\begin{abstract}
We use a versatile model to evaluate the multipartite entanglement and the nonclassical light generation in optical parametric oscillators, exploring the differences between doubly and triply resonant cavity configurations. We demonstrate the entanglement of the pump mode with converted fields in both situations, and the fundamental differences of oscillators using parametric down conversion and four wave mixing processes as the intracavity amplification technique. The strong correlations involving the sidebands of the pump and converted fields gives the signatures of a rich dynamic of multipartite entanglement.
%present a theoretical study about the classical and quantum properties of an optical parametric oscillator (OPO) considering a double and a triply resonant cavities. We show the predictions for two different gain media, $\chi^{(2)}$ and $\chi^{(3)}$. We compare the difference between these OPOs and show that both act as a squeezer source for the pump beam and as a source of entanglement states above the threshold power.
 \end{abstract}
\maketitle
%%%%%%%%%%%%%%%%%%%%%%%% INTRODUCTION %%%%%%%%%%%%%%%%%%%%%%%%%%%%%
% * <bruzaca.rayssa@gmail.com> 2018-09-01T12:24:27.977Z:
\section{Introduction}
Initially presented as a frequency converter that produces tunable coherent radiation \cite{giordmaine1965tunable},  %\cite{burnham1970,friberg1985}. estes dois são OPA/PDC
%,eckardt1986,wong2007,mei2016}.A lista é longa, pois sintonia é todo um domínio na ótica não-linear. Não vale a pena se aprofundar no domínio aqui.
the simplest configuration of an Optical Parametric Oscillator (OPO) was showed to be a source of multiple non-classical states of light. The long list includes squeezed states \cite{Wu1986,Fabre1997}, quantum correlated beams \cite{Heidmann1987, Cassemiro:07}, entangled thermal states \cite{Peng1992}, and multicolor entanglement for two  \cite{Polzik2002, Villar2005}, three  \cite{Coelho823}, and up to six modes of the field \cite{barbosa2018hexapartite}.
This versatile source has found applications in quantum metrology \cite{vahlbruch2007}
%,joo2011 incremental
%,zielinska2014} nenhuma correspondência
and quantum communication protocols, such as quantum key distribution \cite{pereira2000}
%hillery2000,theory
%bennett1994, porzio2007}, superdense coding \cite{bennett1992},
and quantum teleportation \cite{furusawa1998}. Moreover, extensions over the basic configuration allowed the generation of cluster states with arbitrarily large number of modes \cite{furusawa2013,pfister2014} as a possible resource for quantum computation \cite{vanLoock2009}.
% bouwmeester1997,is PDC

The usual configuration consists of a non-linear medium inside an optical cavity. The nonlinear medium will couple the pump field to a pair of modes named signal and idler, often through a second ($\chi^{(2)}$) or third ($\chi^{(3)}$) order non-linearity. Energy exchange among these three fields will follow both energy and momentum conservation \cite{giordmaine1962}.
The nonlinear medium acts as a parametric amplifier and, if amplification matches the cavity losses, the oscillation  threshold is reached and we have the 
generation of intense output fields.
Even for this simple case we may have distinct cavity configurations \cite{Fabre_Cohadon}. The cavity  may be either resonant for all three modes (triply resonant OPO - TROPO), for both converted fields (doubly resonant OPO - DROPO), or just for one of them (singly resonant OPO - SROPO).
% Just as it is possible to have different configurations of cavity, changing the reflections of the cavity mirrors \cite{Laurat:05} or the beams that are resonant in the cavity: triply (TROPO) \cite{barbosa2018hexapartite,MUNOZ2018}, double (DROPO) \cite{Dietrich:19} or single (SROPO) \cite{yang1993,aadhi2017}, showing the OPO versatility.

%As for the gain medium, it is worthy to note that most of the OPOs have a $\chi^{(2)}$ material \cite{phua1999,budni2000}. 

Quantum treatment of a TROPO using a $\chi^{(2)}$ medium was extensively done in both operation regimes (below \cite{graham1968quantum,
%PhysRevA.62.033804, 
Drummond1990} and above threshold \cite{ 
Reid1989,Fabre_1990}). This treatment applies for the  DROPO as well, as far as the pump depletion is negligible. That is the case far below the oscillation threshold,  where pump is treated just as a classical field. To the best of our knowledge, a full quantum  treatment for the DROPO and the effects of pump depletion on the quantum noise is missing. Here we use the multimode analysis of the cavity evolution \cite{MUNOZ2018} to compare the noise and the quantum correlations of DROPO and TROPO in above threshold operation, demonstrating that multipartite entanglement and the squeezing of the pump noise are present even in a single pass regime.
% \cite{PhysRevA.40.913},

The model has the advantage of being applicable to open cavities, beyond the closed cavity regime adopted in typical input/output formalism \cite{yurke1984,yurke1985}. That is particularly interesting for the study of  $\chi^{(3)}$ OPOs, and we compare their behavior to the usual $\chi^{(2)}$ oscillator above threshold.
The small gain, typically found in $\chi^{(2)}$ amplifiers based on   parametric down conversion (PDC), leads to high reflectance mirrors (typically greater than 95\%) for continuous operation even for the best available crystals. On the other hand, although many oscillators based on four wave mixing (4WM)  will present a closed cavity \cite{vahala2004}, 
%Due to the small medium gain, the cavities are made with low transmittance (closed cavities) to increase the re-injection of the fields and enable oscillation above threshold, showing an abrupt threshold transition \cite{phua1999,budni2000}.  
this process can be much stronger in atomic vapors \cite{PhysRevA.78.043816}, where the  $\chi^{(3)}$ process is enhanced close to the atomic transitions.
This enable the development of OPOs with higher transmittance mirrors (open cavity) \cite{yu2010,alvaro2020}. 
%Those open cavities are expected to present a smooth transition through threshold transition. 

%This paper provides a model appropriate to describe the classical behavior of a double resonant OPO and a triple resonant OPO with equal mirrors reflectivities for $\chi^{(2)}$ and $\chi^{(3)}$ media -- suitable to work with both open and closed cavities. We also provide an extension to work with $\chi^{(3)}$ TROPOs with different reflectance coefficients for the pump and signal (idler) beams. 

We begin by a description of the classical behavior of an OPO considering the DROPO and the TROPO case without any approximation for the reflections coefficients of the cavity (section~\ref{ClassAppr}), accounting for the evolution of the mean fields along the gain medium (both $\chi^{(2)}$ and $
\chi^{(3)}$). We go beyond the studies of the classical behavior of $\chi^{(2)}$ \cite{Debuisschert:93} that accounts only for the first order of the expansion to compute the mean fields, which is not appropriate to simulate a $\chi^{(3)}$ OPO with a higher gain \cite{alvaro2020}.
Afterwards, we provide a quantum description of the systems considering the quantum treatment of the fluctuations in terms of the symmetric and antisymmetric basis of the electromagnetic field yielding a full description of the state of the fields in terms of a covariance matrix (section~\ref{Qcorr_sec}). Next, we evaluate the quantum features generated by different types of OPO (section~\ref{secent}), squeezing, bipartite and multipartite entanglement, showing the main differences that appear on each configuration. 
% \cite{PhysRevA.88.052113}

%%%%%%%%%%%%%%%%%%%%%%%%%%%%%%%%%%%%%% CLASSICAL APPROACH %%%%%%%%%%%%%%%%%%%%%%%%%%%%%%%%%%%
\section{Classical Approach}\label{ClassAppr}
We evaluate the evolution of the  mean field during the parametric amplification in both cases, PDC and 4WM, before considering the steady state conditions inside a cavity. The cavity feedback will lead to a dramatic effect of gain saturation, completely modifying the response of the free propagating process.

Using a medium with second (third) order non-linearity, one (two) photon(s) of the pump beam (of frequency $\omega_1$) might be converted into two photons, signal and idler (with respective frequencies $\omega_1$ and $\omega_2$). The interaction Hamiltonians $\hat{H}^{{(2)}}$, which represents the PDC, and $\hat{H}^{{(3)}}$, representing the 4WM, are given by:
\begin{eqnarray}
\hat{H}^{{(2)}}&=&i\hbar \chi^{(2)}\hat{a}_{0}(t)\hat{a}_{1}^{\dagger}(t)\hat{a}_{2}^{\dagger}(t)-h.c.,
\label{hintchi2}
\\
\hat{H}^{{(3)}}&=&i\hbar\chi^{(3)} \hat{a}_0^2(t) \hat{a}_1^{\dagger}  (t)\hat{a}_2^{\dagger}(t) + h.c.,
\label{HamiltonianoInteracaoTemporal}
\end{eqnarray} 
in the interaction picture. Here $\hat{a}_n$, with $n=\{0,1,2\}$, represents the annihilation operators of the pump, signal and idler modes, respectively, and the parameter $\chi^{(m)}$ with $m=\{2,3\}$ is associated with the non-linear susceptibility coefficient of each gain medium. 

In order to determine the mean amplitudes of the output fields as a function of the system parameters, the operators evolution $\hat{a}_n(t)$ through the gain medium are evaluated through the Heisenberg equations for the field operators, $\frac{d}{d t}\hat{a}_{n}(t)=(i/\hbar)[ \hat{H}^{(m)},\hat{a}_{n}(t)]$. 
Linearizing the field operators as $\hat{a}_{n}(t) = \alpha_{n} + \delta\hat{a}_{n}(t)$, where $\alpha_{n}$ is the mean field amplitude and $\delta\hat{a}_{n}(t)$ is the field fluctuations, the set of expressions that describe the mean value evolution of the fields $\alpha_{n}$ through the gain medium can be written as: 
\begin{subequations}
\label{eq_heisemberg_amplitudes1}
\begin{align}
  \frac{d \alpha_{0}}{d t} &= -(m-1)\chi^{(m)}\alpha_{0}^{*(m-2)}\alpha_{1}\alpha_{2}, \label{eq_HA_1} \\ 
\frac{d \alpha_{1}}{d t} &= \chi^{(m)}\alpha_{0}^{(m-1)}\alpha_{2}^{*}, \label{eq_HA_2} \\
\frac{d \alpha_{2}}{d t} &= \chi^{(m)}\alpha_{0}^{(m-1)}\alpha_{1}^{*}, \label{eq_HA_3}
\end{align}
\end{subequations}
where $\alpha^{*}_{n}$ is the complex conjugate of $\alpha_{n}$. The mean value of the field is a complex number that can be explicitly written in terms of real amplitude and phase: $\alpha_{n}=\sqrt{P_n}e^{i \theta_n}$. 
The parameter $P_n=\alpha_{n}\alpha_{n}^*$ is proportional to the photon number in the field $n$ and hence to the field power. 
Differentiating $P_n$ in time we have
\begin{align}
\frac{d}{dt}P_n&= \alpha_n \frac{d \alpha_n^*}{dt}+\alpha_n^*\frac{d \alpha_n}{dt},
\label{power}
\end{align}
and with the help of Eq.(\ref{eq_heisemberg_amplitudes1}),
we have a set of differential equations that describe the evolution of $P_n$%, we have:
\begin{align}
\label{difpower}
\frac{d P_0}{dt} &= -2(m-1) \chi^{(m)}
\sqrt{P_{0}^{(m-1)}P_1 P_2} \,\cos{\theta}, \nonumber  \\
\frac{d P_1}{dt} &= 2 \chi^{(m)} \sqrt{P_{0}^{(m-1)}P_1 P_2} \,\cos{\theta}, \nonumber \\ 
\frac{d P_2}{dt} &= 2 \chi^{(m)} \sqrt{P_{0}^{(m-1)}P_1 P_2} \,\cos{\theta},
\end{align}
depending on a global phase $\theta= \theta_1 + \theta_2-(m-1)\theta_0$.  In practice, the exact value will depend on the phase matching condition, involving the value of $\chi^{(m)}$ and on the reflection coefficient of the mirrors \cite{Debuisschert:93}, but the general effect will be the modulation of the coupling. Therefore, we will chose $\theta$ that maximizes the coupling, thus $\cos(\theta)= 1$.
A detailed evaluation of the field evolution for the case of a $\chi^{(2)}$ medium is given in \cite{rosencher2002}. 

As the fields propagate along the gain medium, we may evaluate the power transfer from the pump to the converted modes. 
The number of photons of the pump will be reduced, leading to $P_0(t)=P_0(0)-(m-1)p(t)$. This photon depletion leads to a change in the photon number of the converted fields as $P_1(t)=P_1(0)+p(t)$ and $P_2(t)=P_2(0)+p(t)(t)$.
Considering a balanced power for converted modes, $P_1(t)=P_2(t)$, therefore the set of Eqs. (\ref{difpower}) can be used to obtain the evolution of $p(t)$, related with the power transfer along the path. For convenience, 
\begin{equation}  \label{eq_dx/dt}
\frac{dp}{dt}=2\chi^{(m)}(P_0(0)-(m-1)p)^{\frac{1}{2}(m-1)}(P_1(0)+p).
\end{equation}\par

The total power variation of signal field, $\Delta P_{1}^{(m)}$, for each gain medium $m$, will be computed integrating  Eq. (\ref{eq_dx/dt}) in the limits of the entrance $(t_{1},p(t_{1}))=(0,0)$, and the end $(t_2,p(t_2)) = (L/(\nn c), \Delta P_{1}^{(m)})$ of the gain medium of length $L$, where $c$ is the velocity of light in the vacuum and $\nn$ is the refractive  index. 
Considering the $\chi^{(2)}$ gain medium we have
\begin{eqnarray}\label{gainchi2}
\Delta P_{1}^{(2)}&=& P_0(0)-(P_0(0)+P_1(0)) \nonumber \\
&\times& tanh^2 \left[\kappa^{(2)} \sqrt{P_0(0)+P_1(0)} \right.  \nonumber \\ &-& arctanh \left.\left[ \sqrt{\frac{P_0(0)}{P_0(0)+P_1(0)}}\right]\right],
\end{eqnarray}
with $\kappa^{(m)} = \chi^{(m)} L/(\nn c)$. Although this equation is not simple, in the limit of a weak coupling we have $\Delta P_1^{(2)} =  \kappa^{(2)} /P_1(0)\sqrt{P_{0}}$, recovering the situation observed in \cite{Debuisschert:93}.  
%is a constant proportional to the interaction time of the photons with the gain medium. 
%Trocar nesse caso G por G. Rever esse parágrafo.
%The relative gain using the $\chi^{(2)}$ gain medium, $G_{2}=\frac{\Delta P1}{P1}=\kappa_{2}\sqrt{P_{0}}$, increases linearly with the pump power and with the square root of $\kappa_{2}$. Considering a common situation, where $G_{2} \approx 4\%$ and the pump power intracavity is computed following \cite{Debuisschert:93}, an estimated value of $\kappa_{2}$ can be obtained. 
%In this paper we will set $\kappa_{2}=0.5 W^{-1}$ in all the simulations using the $\chi^{(2)}$ gain medium in order to compare the main results using this system with the $\chi^{(3)}$ OPO results.
For a $\chi^{(3)}$ gain medium we obtain
\begin{equation}\label{resultado_ganho_1}
\Delta P_{1}^{(3)} = \frac{P_0(0) P_1(0) \left(e^{2 \kappa^{(3)} (P_0(0)+2P_1(0))}-1\right)}{2 P_1(0) e^{2 \kappa^{(3)} (P_0(0)+2 P_1(0))}+P_0(0)}.
\end{equation}

In the $\chi^{(3)}$ scenario, it is interesting  to evaluate the signal (and idler) relative gain $G^{(3)}$. From  Eq.  (\ref{resultado_ganho_1}) we have
\begin{equation}\label{resultado_ganho_2}
G^{(3)}=\frac{\Delta P_{1}^{(3)}}{P_1(0)} =\frac{ g^{(3)}-1}{1+2 g^{(3)} P_1(0)/P_0(0)},
\end{equation}
where  $g^{(3)}=\exp{[2 \kappa^{(3)} (P_0(0)+2P_1(0))]}$ corresponds to the unsaturated amplification. For a completely open cavity and a weak seed ($P_1(0)\ll P_0(0)$),  Eq. (\ref{resultado_ganho_2})  simplifies as
\begin{equation}\label{resultado_ganho_3}
G^{(3)}+1=e^{2 \kappa^{(3)} P_0(0)}=g^{(3)}.
\end{equation}
The amplification then increases exponentially with the pump power and the medium length, which is proportional to $\kappa^{(3)}$. This is the typical situation of unsaturated parametric amplifier as used in \cite{PhysRevA.78.043816}. 
It also becomes evident the role of an increasing seed: it will lead to a saturation of the power transfer process, and therefore a reduction of the gain in Eq. (\ref{resultado_ganho_2}).

From Eq. (\ref{gainchi2}) and (\ref{resultado_ganho_1}), we can relate the added power on the converted fields to the coupling coefficients $\kappa^{(m)}$, and the field power at the input of the amplifier, $P_0(0)$ and $P_1(0)$. This result is used now to evaluate the intracavity steady state. 
First, we compare the solutions for both gain media in a doubly resonant cavity in Section \ref{DROPO_R}, where pump beam makes a single pass through the cavity while signal and idler beams are resonant. 
Next, in Section \ref{TROPO_R}, we consider a triply resonant OPO, exploring different reflections of the input mirror for the pump field.
%with the same reflection coefficient for all three resonant fields, a typical situation for a $\chi^{(3)}$ medium. In the last subsection, \ref{TROPO_Rdif}, we study the triply resonant cavity considering that reflection coefficients for the pump and converted beams are different.

\subsection{Doubly Resonant Optical Parametrical Oscillator}\label{DROPO_R}
\begin{figure}[t]
	\begin{center}
		\includegraphics[scale=1.3]{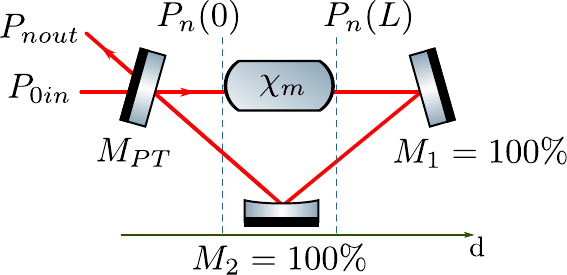}
		\caption{OPO model using a ring cavity with two highly reflective mirrors ($M_{1}=M_{2}=1$), an output coupler $M_{PT}$ and a gain medium with length $L$. The $M_{PT}$ reflectante coefficient for the pump beam is $R_{0}$ and for both, signal and idler beams, $R_{1}$.}
 		\label{DROPO}
	\end{center}
\end{figure}

The first system being studied is a DROPO (Fig. \ref{DROPO}). For convenience, we will label the power at the input and the output of the medium by the positions $0$ and $L$.
In this case, the pump beam makes a single pass through the cavity, therefore the pump power injected in the cavity defines $P_{0in}=P_0(0)$. When it leaves the cavity, the output power %$P_{0out}=P_0(L)$ 
is given by
%\begin{eqnarray}
$P_{0out}=P_0(L)=P_{0in}-(m-1)\Delta P_{1}^{(m)}$.
%\end{eqnarray}

In the steady state, we can relate the input power of the signal on the amplifier to the output of the amplifier using the reflectance $R_{1}$ on the output coupler $M_{PT}$, as $P_1(0)=R_{1}P_1(L)$. The output of the OPO will be related to the output of the amplifier as well $P_{1out}=(1-R_{1})P_1(L)$. From the definition of $\Delta P_1^{(m)}=P_1(L)-P_1(0)$
%=(1-R_{1})(P_1(0)+\Delta P_1^{(m)})$.
%=R_{1}(P_1(0)+\Delta P_1^{(m)})
%In the steady state, we can relateConsidering the system in equilibrium, in a complete round trip the  beam $P_{1}(0)$  at the gain medium input is the result of the re-injection of the reflected portion of $P_{1}(L)$ in the mirror $M_{PT}$ with reflectance $R_{1}$ thus $P_{1}(0)=R_{1}P_{1}(L)$. After passing through the gain medium, the signal (and idler) power is $P_{1}(L)=P_{1}(0)+ G^{(m)}$ and the output field is given by:
%when we can see the effect of the non-linear conversion term. 
we can calculate the output power $P_{1out}$ as
\begin{equation}
P_{1out}=\frac{(1-R_{1})P_{1}(0)}{R_{1}}= \Delta P_{1}^{(m)}.
\label{eq_P_1out}
\end{equation}
%The same behavior is expected in the signal power, $P_2^{out}$. 
This result is already expected from a cavity in equilibrium due to energy conservation: the energy added to a given mode will match the losses through the output coupler.

%%%%%%%%%%%%%%%%%%%%%%%%%%%%%%%%%%%%%%% FIGURE DROPO %%%%%%%%%%%%%%%%%%%%%%%%%%%%%%%%%%%%%%%%%%%%%%%
%%%%%%%%%%%%%%%%%%%%%%%%%%%%%%%%%%%%%%%%%%%%%%% INSIDE CAVITY %%%%%%%%%%$%%%%%%%%%%%%%%%%%%%%%%%%%%%%

\begin{figure}[ht]
	\begin{center}
		\includegraphics[scale=0.3]{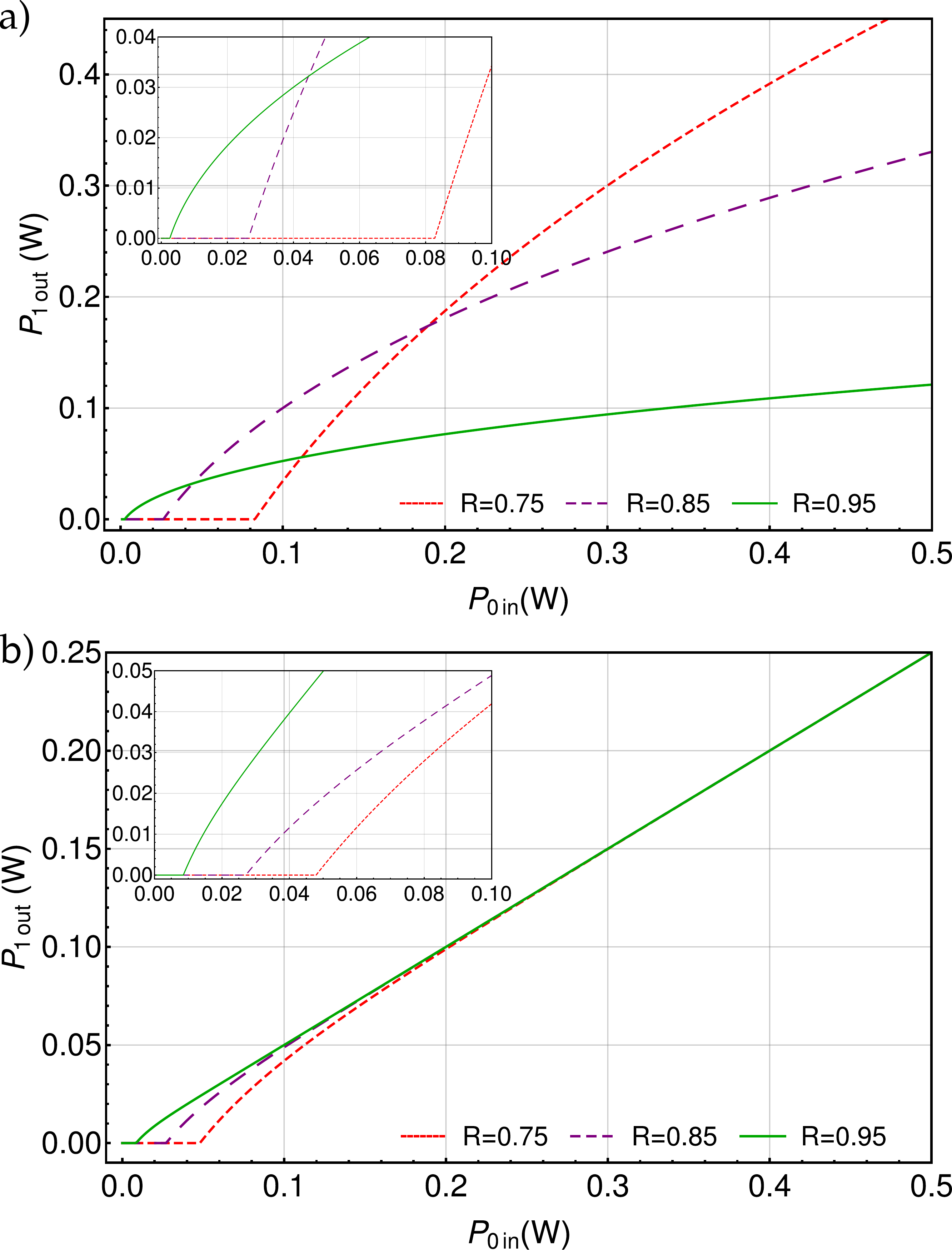}
		\caption{DROPO: $P_{1out}$ as a function of $P_{0in}$ for different reflectivity coefficients $R_{1} = \{75\%, 85\%, 95\%\}$. Simulations considering (a) a $\chi^{(2)}$ gain medium with $\kappa^{(2)}=0.5 W^{-1/2}$ and in (b) a $\chi^{(3)}$ gain medium with $\kappa^{(3)}=3 W^{-1}$.}
		\label{chi2chi3pth}
	\end{center}
\end{figure}

%This section main result is the behavior of the signal (idler) output power in terms of the input pump power, $P_{0in}$. 
Numerical solution of Eq.~ (\ref{eq_P_1out}), combined with ~Eq. (\ref{gainchi2}) (for $\chi^{(2)}$) and~Eq. (\ref{resultado_ganho_1}) (for $\chi^{(3)}$)  gives the value of $P_1(0)$ as a function of the pump power $P_{0in}$. Evaluation of the output power  $P_{1out}$ is immediate. The result is presented in
%The output power is zero below the threshold power. 
%From this point, it is necessary to evaluate separately the behavior of the generated output fields for each medium. 
Fig. ~\ref{chi2chi3pth}(a), for the  $\chi^{(2)}$ gain medium, giving an output power close to the parabolic curve deduced in \cite{rosencher2002}. On the other hand, in Fig.~\ref{chi2chi3pth}(b), for the  $\chi^{(3)}$ gain medium, we can observe that the curve approaches a proportional response for sufficiently high pump power. Even close to the threshold, as showed in the inset, the evolution could be closely approximated by a linear response when the cavity coupling is higher, as observed in \cite{alvaro2020}. 
The asymptotic behavior is similar for the distinct couplings, and very different from the one observed with  $\chi^{(2)}$.
The insets put in evidence the reduction of the threshold power for a reduction of the cavity losses. 
The chosen value of the nonlinearity  $\kappa^{(3)}=3 W^{-1}$ is based on the observed amplification in  \cite{PhysRevA.78.043816}. The value of $\kappa^{(2)}=0.5 W^{-1/2}$ is chosen to match the threshold power for both media with $R_1=0.85$. We will adopt this value for the simulations along the remaining of the article.

\begin{figure}[ht]
\centering
\includegraphics[scale=0.3]{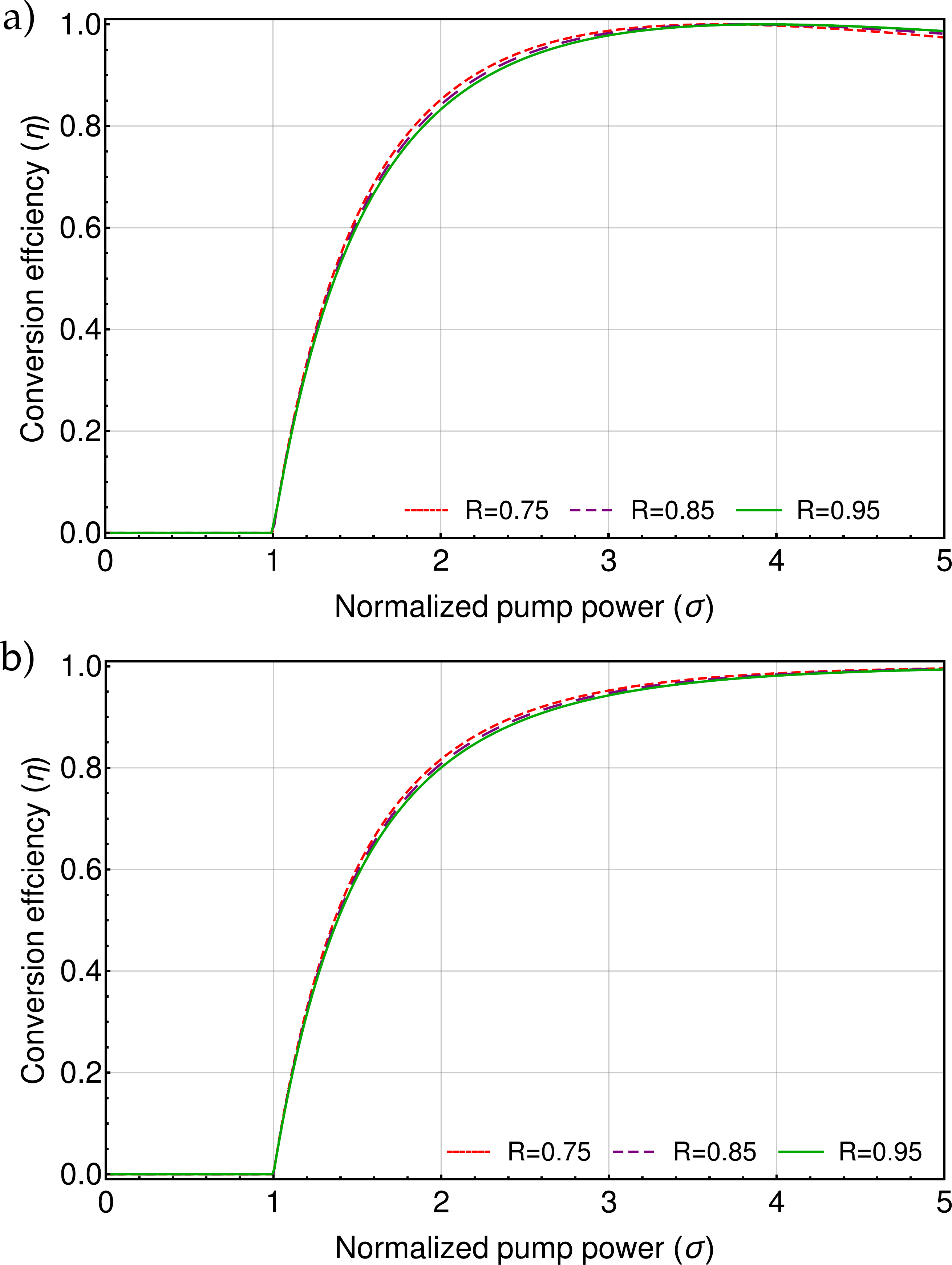}
\caption{\label{chi2chi3ef} DROPO: Conversion efficiency $\eta$ as a function of pump power (normalized by the threshold power) $\sigma$ for $R_{1} = \{75\%, 85\%, 95\%\}$. Simulations with a) a $\chi^{(2)}$ gain medium and b) a $\chi^{(3)}$ gain medium.}
 \end{figure}

Although the power has some dramatic changes for distinct coupling, a better comparison can be done using the conversion efficiency 
\begin{equation}
\eta=\frac{\hbar\omega_1 P_{1out}+\hbar\omega_2 P_{2out}}{\hbar\omega_0 P_{0in}}.
\end{equation}
The efficiency is showed in Fig.~\ref{chi2chi3ef}, with the pump power normalized to the threshold power. The conversion efficiency increases monotonically in both cases, but the most relevant difference between $\chi^{(2)}$ and $\chi^{(3)}$ amplifiers comes from the fact that a maximum (unitary) efficiency is observed at a pump power $\simeq 4$ times above threshold for $\chi^{(2)}$ oscillators, while $\chi^{(3)}$ oscillators evolve asymptotically to unitary gain. Notice as well that while the unitary efficiency for $\chi^{(2)}$ is reached at $\sigma_{max} \simeq 4$ for a closed cavity, the position of the maximum is reduced when cavity losses are increased, as observed in \cite{rosencher2002}.

%is observed with the input pump power reaching the maximum approximately 4 times above the threshold power in both cases. After this point, the efficiency behavior in a system with a~$\chi^{(2)}$ gain medium decreases in function of the input pump power  due to the depletion of the input pump beam (\autoref{chi2chi3ef}(a)). Using a $\chi^{(3)}$ gain medium, after reaching the maximum efficiency the system enters in a stationary state, in which approximately all the pump power will be converted in signal and idler beams (\autoref{chi2chi3ef}(b)).

%%%% OUTPUT POWER
%Mencionar no texto os inset dos gráficos
%%%%
%%%%%%%%%%%%%%%%%%%%%%%%%%%%%%%%%%Output power %%%%%%%%%%%%%%%%%%%%%%%%%%%%%%%%%%%%%%%%%%%%%%%%%%%%%%%%
%%%%%%%%%%%%%%%%%%%%% Efficiency Conversion %%%%%%%%%%%%%%%%%%%%%%%%%%%%%%%%%%%%%%%%%%%%%%%%%%%%%
 
%%%%%%%%%%%%%%%%%%%%%%%%%%%%%%%%% Conversion Efficiency %%%%%%%%%%%%%%%%%%%%%%%%%%%%%%%%%%%%%%%%%%%%%%%%%%%%%%%%

 %%%%%%%%%%%%%%%%%%%%%%%%%%%%%%%%%% TROPO %%%%%%%%%%%%%%%%%%%%%%%%%%%%%%%%%%%%%%%%%%%%%%%%%%%%%%%%
\subsection{Triply Resonant Optical Parametrical Oscillator}\label{TROPO_R}
%In the triply resonant configuration, pump, signal and idler fields are resonant in the cavity. 
We have now a cavity for the pump that enhances its power, what makes this system being called pump enhanced DROPO as well.
Self-consistency equation for the converted field, Eq. (\ref{eq_P_1out}), is still valid.
However the self consistency equations will differ for the pump field that now includes a beam splitter transformation for the input coupler:
\begin{align}
\sqrt{P_{0}(0)}&= t_0\sqrt{P_{0in}} - r_0 \sqrt{P_{0}(L)} \label{eq:sqrtP00}\\
\sqrt{P_{0out}}&= t_0\sqrt{P_{0}(L)} + r_0 \sqrt{P_{0in}}\label{eq:sqrtP0out},
\end{align}
where $r_0=\sqrt{R_0}$ is the reflection coefficient and $t_0=\sqrt{1-R_0}=\sqrt{T_0}$ is the transmission coefficient.

The output pump power in terms of the input pump power $P_{0in}$ and the intracavity pump power $P_{0}(0)$ is obtained by combining Eqs. (\ref{eq:sqrtP00}) and (\ref{eq:sqrtP0out}):
\begin{equation}\label{eq:P0out}
P_{0out}=\frac{P_{0in}-2t_0 \sqrt{P_{0in}P_{0}(0)}+T_0 P_0(0)}{R_0}.
\end{equation}

We would like to evaluate the behavior of $P_{1out}$ in terms of $P_{0in}$. A numerical solution for this quantity can be obtained by evaluating the intracavity fields $P_{0}(0)$ and $P_{1}(0)$ as a function of the input power, the coupling constant $\kappa^{(m)}$ and the mirror reflectances $R_0$, $R_1=R_2$. As an example, a detailed evaluation of the output power for $\chi^{(3)}$ TROPO is performed in the Appendix \ref{App_TROPORdif}.

%A numerical solution can now be obtained from Eqs. (\ref{eq_P_1out}) and (\ref{eq:P0out}) for the cavity, and Eqs. (\ref{gainchi2}) and (\ref{resultado_ganho_1}) for the gain, in order to evaluate $P_{0}(0)$ and $P_{1}(0)$ as a function of the input power $P_{0in}$, the coupling constant $\kappa^{(m)}$ and the mirror reflectances $R_0$, $R_1=R_2$. As an example, a detailed evaluation of the output power for $\chi^{(3)}$ TROPO is performed in the Appendix \ref{App_TROPORdif}.

%Note that the field evolution through the non-linear media does not change with the cavity configuration. Thus,  Eqs. (\ref{gainchi2}) and (\ref{resultado_ganho_1}) are still valid in the TROPO case due to their exclusive dependence on input field powers $P_{0}(0)$, $P_{1}(0)$ and the coupling constant $\kappa_{m}$.

For the moment we will consider that the cavity mirrors have the same reflectance for pump, signal and idler beams, $R_0=R_1=R_2=R$. While this situation is quite unusual for $\chi^{(2)}$ OPOs, it is common for the $\chi^{(3)}$ condition, when all the resonant fields may be nearly degenerate in frequency, leading to balanced losses. In this situation a direct equation for the intracavity fields and the pump power can be obtained.

In a steady state the total intracavity power, $P_{T0}$, is constant at any point inside the cavity, resulting in the relation $P_{T0} =P_{0}+(m-1)P_{1}$. Due to energy conservation, the relation between the input pump field and the output fields of the cavity is $P_{0in}=P_{0out}+(m-1)P_{1out}$, where $P_{1out}$ is given by  Eq. (\ref{eq_P_1out}). Now it is possible to rewrite the intracavity fields, $P_{0}(0)$ and $P_{1}(0)$,  in terms of the input pump power $P_{0in}$, and the total intracavity pump power, $P_{T0}$. From the above description, combined  %=P_{0in}-(m-1)(1-R)P_{1}(0)/R$ 
with  Eqs. (\ref{eq:sqrtP00}) %results in $P_{0}(0)$. Then, 
%remembering that $P_{0}(L)$ is the result of the pump field propagation through the gain medium %, $P_{0}(L)=P_{0}(0)-(m-1)G=P_{0}(0)-(m-1) (1-R)P_{1}(0)/R$
%and replacing it in  Eq. 
and (\ref{eq:P0out}), we obtain % $P_{1}(0)$, we can obtain for each case. The final expression for each case is:
\begin{eqnarray}
	P_{0}(0)&=&\frac{1-R}{4}\frac{(P_{0in}+P_{T0})^2}{P_{0in}} \label{eq:P00},  \\
		P_{1}(0)&=& \frac{R-1}{4(m-1)}\frac{(P_{0in}-P_{T0})^2}{P_{0in}}+\frac{RP_{T0}}{m-1}\label{eq:P10},
\end{eqnarray}
where  Eq. (\ref{eq:P00}) and  Eq. (\ref{eq:P10}) describe the behavior of $P_{0}(0)$  and $P_{1}(0)$ as a function of $P_{T0}$ and $P_{0in}$.

%%%%%%%%%%%%%%%%%%%%%%%%%%%%%%%%%%% FIGURE Mean field %%%%%%%%%%%%%%%%%%%%%%%%%%%%%%%%%%%%%%%%%%%%%%%%%%
 \begin{figure}[t]
	\centering
	\includegraphics[scale=0.3]{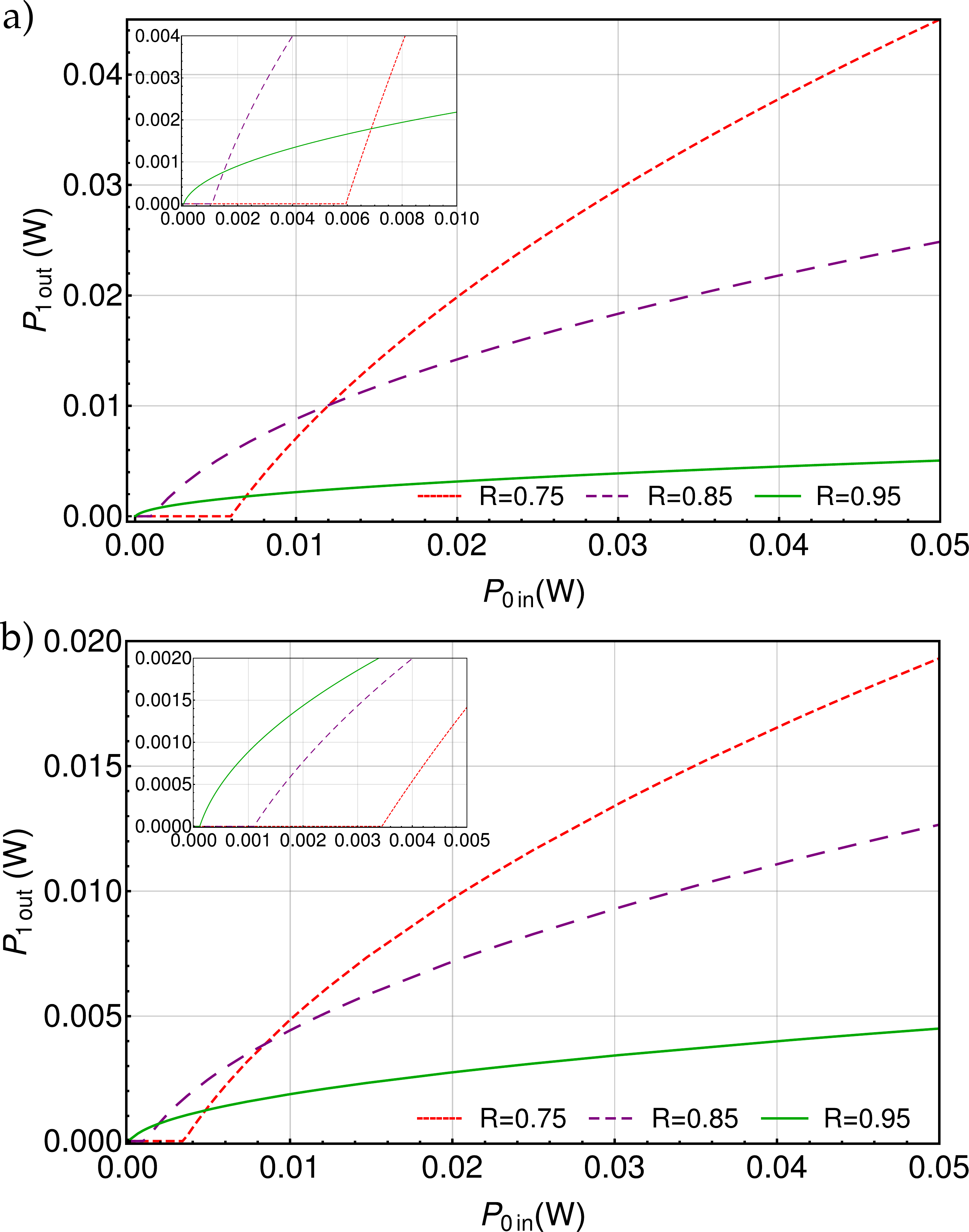}
 \caption{\label{fig:TROPOMEDIO} TROPO: $P_{1out}$ in function of $P_{0in}$ for different reflectivity coefficients $R_{1} = \{75\%, 85\%, 95\%\}$. Simulations considering a) a $\chi^{(2)}$ gain medium and in b) a $\chi^{(3)}$ gain medium.}
\end{figure} 

 \begin{figure}[t]
	\centering
	\includegraphics[scale=0.3]{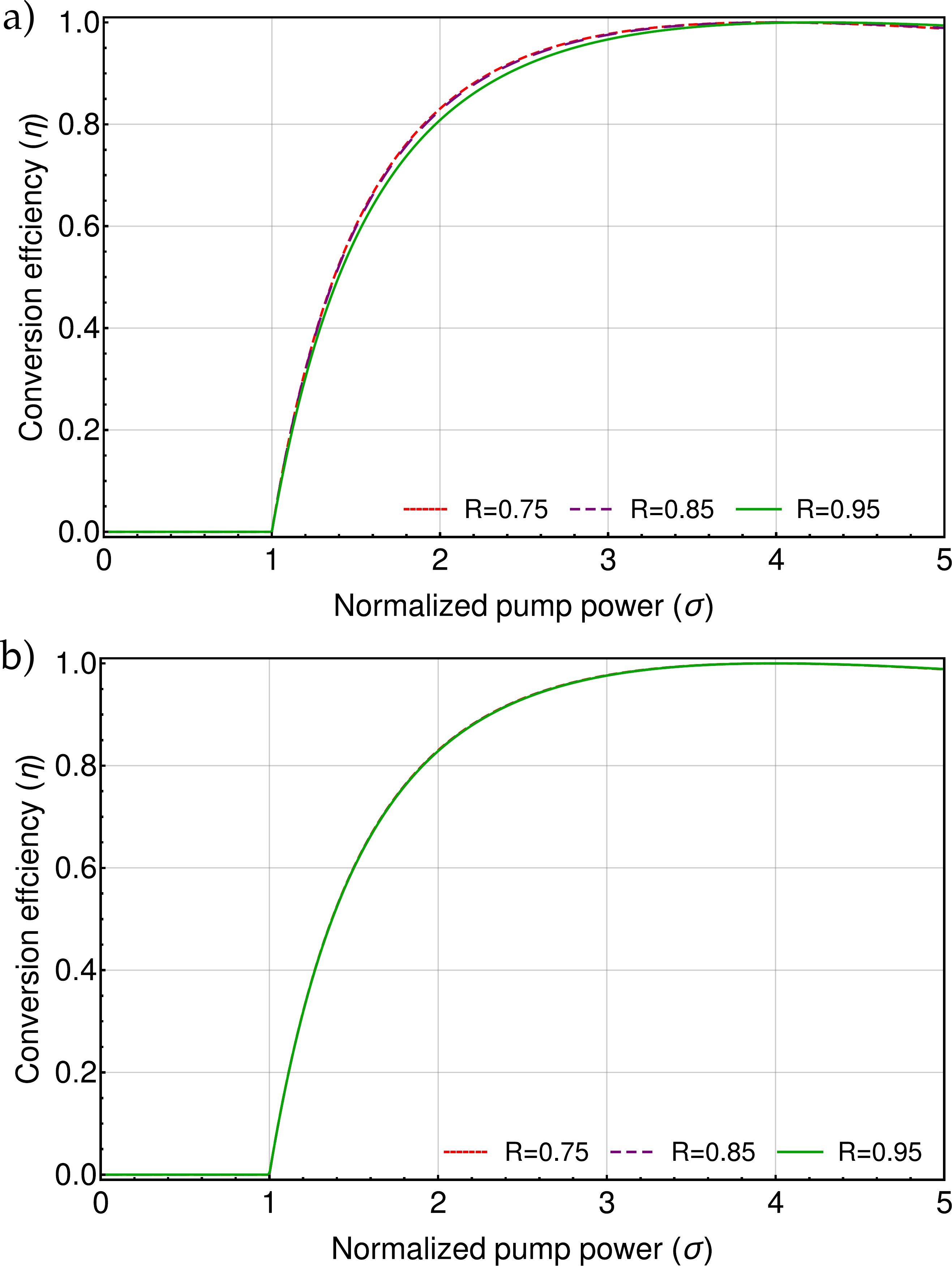}
 \caption{\label{fig:TROPOQE}TROPO: $\eta$ in function of $\sigma$ for $R_{1} = \{75\%, 85\%, 95\%\}$. Simulations with a) a $\chi^{(2)}$ gain medium and b) a $\chi^{(3)}$ gain medium.}
\end{figure} 

%The problem now is to write $P_{T0}$ as a function of the input pump power, $P_{0in}$. In the TROPO  $\chi^{(2)}$ configuration, we equally Eq. 7 e Eq. 11 and in TROPO  $\chi^{(3)}$ we equally  Eq. 8 Eq 11, then we substitute the Eqs. 15 and 16 on those expressions to find a numerical solution for $P_{T0}$.

The problem now becomes writing $P_{T0}$ as a function of the input pump power, $P_{0in}$. In the $\chi^{(2)}$ TROPO  configuration, by equalling  Eq. (\ref{gainchi2}) and Eq. (\ref{eq_P_1out}) results in
\begin{align}\label{eq:PT0Pin2}
\frac{(1-R)(P_{0in}-P_{T0})^2}{4 R P_{0in} P_{T0}} =& 
tanh^2 \left[\kappa^{(2)} \sqrt{P_{T0}} \right.\\
- arctanh &\left.\left[ \sqrt{(1-R)\frac{(P_{0in}+P_{T0})^2}{4P_{0in}P_{T0}}}\right]\right],  \nonumber 
\end{align}
which is used to find a numerical solution to $P_{T0}$ as a function of $P_{0in}$. 
Considering the $\chi^{(3)}$ TROPO, the treatment consists of equalling  Eq. (\ref{resultado_ganho_1})  to Eq. (\ref{eq_P_1out}). Rearranging the terms we obtain $\frac{P_1(0)}{P_0(0)} = \frac{R e^{2\kappa_{3} P_{T0}}-1}{2(1-R) e^{2\kappa_{3} P_{T0}}} $
and substituting   Eq. (\ref{eq:P00}) and  Eq. (\ref{eq:P10}) in the later expression results in:
% $\frac{P_1(0)}{P_0(0)} = \frac{R e^{2\kappa_{3} P_{T0}}-1}{2(1-R) e^{2\kappa_{3} P_{T0}}} $
\begin{align}\label{eq:PT0Pin3}
(1-R)(P_{0in}-P_{T0})^2 &- 4 R P_{0in}P_{T0}\\ \nonumber
 =&(e^{-2\kappa_{3} P_{T0}}-R)(P_{0in}+P_{T0})^2,
\end{align}
enabling a numerical solution for $P_{T0}$ as a function of $P_{0in}$. 
Using equations  Eq. (\ref{eq:PT0Pin2}) and  Eq. (\ref{eq:PT0Pin3}) as an input in equation  Eq. (\ref{eq:P10}) we have $P_{1}(0)$ as a function of $P_{0in}$ for $\chi^{(2)}$ and $\chi^{(3)}$ TROPO, respectively. In order to evaluate the behavior of $P_{1out}$ as a function of $P_{0in}$ we just replace $P_{1}(0)$ in  Eq. (\ref{eq_P_1out}).

The behavior of $P_{1out}$ as a function of $P_{0in}$ is showed for a $\chi^{(2)}$ gain medium in Fig.~\ref{fig:TROPOMEDIO}~(a) and  for a $\chi^{(3)}$ gain medium in Fig.~\ref{fig:TROPOMEDIO}~(b). The behavior of both systems are analysed for three different reflection coefficients and the increase in the threshold power can be observed when $R$ changes from $R=95\%$ to $R=75\%$. Furthermore, the converted fields power increases with the input pump power, but differently from the DROPO, both curves present a parabolic-like shape. 
%The conversion efficiency behavior is showed in the appendix \autoref{App_QE}.

%\section{Quantum efficiency}\label{App_QE}
%%%%%%%%%%%%%%%%%%%%%%%%%%%%%%%%%%% FIGURE QE %%%%%%%%%%%%%%%%%%%%%%%%%%%%%%%%%%%%%%%%%%%%%%%%%%
A better comparison of the curves can be obtained from the conversion efficiency, as showed in Fig.  \ref{fig:TROPOQE}. The maximum conversion efficiency occurs approximately in $\sigma=4$ decreasing from this point on to approximately 50\% for $\sigma \gg 4$. The behavior for the different reflectivity coefficient is very similar in all $\sigma$ analysed for both  $\chi^{(2)}$ and $\chi^{(3)}$ TROPO in the presented range.

These results are similar to those obtained for $\chi^{(2)}$ OPOs, as presented in \cite{Debuisschert:93,rosencher2002}. The main point of the present treatment is to obtain a detailed evolution of the mean fields inside the gain medium. That is a fundamental part in the evaluation of the noise for a cavity in the open limit, beyond the first order approximation for the mean field, as we demonstrate now.

%%%%%%%%%%%%%%%%%%%%%%%%%%%%%%%%%%%%%%%%%%%%% QUANTUM CORRELATIONS %%%%%%%%%%%%%%%%%%%%%%%%%%%%%%%%%%%%%%%%%%
\section{Theoretical description of the field quantum fluctuations}\label{Qcorr_sec}

In order to evaluate the quantum fluctuations as a function of the oscillator parameters in the spectral domain, 
%we need to first describe the OPO Hamiltonian in the sideband space. Following 
we follow the formalism described in~\cite{MUNOZ2018}. We begin by writing the time-dependent annihilation operator in terms of the annihilation operators acting on the modes of the sideband frequencies of the central carrier frequency $\omega_n$ %\cite{walls2007quantum}: 
\begin{eqnarray}
\hat{a}_{n}(t)&=&\int_{-\omega_{n}}^{\infty} e^{-i\Omega t} \hat{a}_{\Omega}d\Omega,
\label{eq_TFBL}
\end{eqnarray}
where $\hat{a}_{\Omega}$ is the photon annihilation operator in the mode of frequency $\Omega=(\omega-\omega_{n})$, $\Omega$ represents the sideband frequency and $\omega_{n}$ is the carrier frequency of mode $n$. Considering a narrow optical spectra for the carrier, $\omega_{n}\gg |\Omega|$, the integral limit can be approximated as $\omega_{n} \rightarrow \infty$.
In the linearized form, the annihilation operator is 
%$ \hat{a}_{\pm\Omega}^{(n)}=\langle\hat{a}^{(n)}_{\pm\Omega}\rangle + \delta\hat{a}_{\pm \Omega}^{(n)}$, where $\pm \Omega$ is a short notation for $\omega_{n}\pm \Omega$. In these terms,  Eq. (\ref{eq_TFBL}) 
can be rewritten as
\begin{eqnarray}
\hat{a}_{n}(t)=\langle\hat{a}_{n}\rangle + \delta\hat{a}_{n}(t)= \alpha_{ n}+ \int^{'} e^{-i\Omega t}\hat{a}_{\Omega ,n}d\Omega,
\label{mn}
\end{eqnarray}
where $\alpha_{n}$ represents the mean value of the carrier at frequency $\omega_n$ and the symbol $^{'}$ in the integral represents the integration limits between $-\infty$ to $\infty$ relative to sidebands frequency disregarding the carrier term. This integral will give rise to the fluctuation term $\delta \hat a_n(t)$.

The interaction Hamiltonian of each interaction process,  Eq. (\ref{hintchi2}) and~(\ref{HamiltonianoInteracaoTemporal}), can now be rewritten with the help of  Eq.~(\ref{mn}). Higher order terms in fluctuation and rapidly oscillating terms, that don't satisfy energy conservation, are neglected. The constant part having only the mean fields is removed as well. We are left only with the contributions of the sidebands, given by
%terms that does not satisfy phase-matching condition are neglected. 
\begin{eqnarray} 
\hat H^{(m)}=\int_{\epsilon}^{\infty} d \Omega \hat H^{(m)}(\Omega) 
\label{hbl1}
\end{eqnarray}
where the sum is taken form a lower frequency component $\epsilon$ defined by the bandwidth of the pump field. The contributions of the sidebands add linearly, and are described by
\begin{eqnarray} 
\hat H^{(m)}(\Omega)&=&i\hbar\chi^{(m)} \Omega[\alpha_{{0}}^{*(m-1)}(\hat{a}_{\Omega,1} \hat{a}_{-\Omega,2}+\hat{a}_{-\Omega,1} \hat{a}_{\Omega,2}) \nonumber \\
&+&(m-1)\alpha_{0}^{*(m-2)}\alpha_{1}(\hat{a}^{\dagger}_{\Omega,0}\hat{a}_{\Omega,2}+\hat{a}^{\dagger}_{-\Omega,0}
\hat{a}_{-\Omega,2})\nonumber \\
&+&(m-1)\alpha_{0}^{*(m-2)}\alpha_{2}(\hat{a}^{\dagger}_{\Omega,0}\hat{a}_{\Omega,1}+\hat{a}^{\dagger}_{-\Omega,0}
\hat{a}_{-\Omega,1}) \nonumber \\
&+&(m-2)\alpha_{1}\alpha_{2}(\hat{a}^{\dagger}_{\Omega,0}\hat{a}^{\dagger}_{-\Omega,0}+\hat{a}^{\dagger}_{-\Omega,0}\hat{a}^{\dagger}_{\Omega,0})-h.c.]. \nonumber \\
\label{hbl2}
\end{eqnarray}

A convenient form of writing this Hamiltonian is using a 
%is convenient to rewrite the interaction Hamiltonian in terms of 
symmetric and antisymmetric combination of modes, defined as: 
$\hat{a}_{ns/a}=[\hat{a}_{\Omega,{n}} \pm \hat{a}_{-\Omega,{n}}]/\sqrt{2}$, where $+~(-)$ signal refers to symmetric (antisymmetric) sideband operators thus simplifying the Hamiltonian into two terms, $\hat H^{(m)}(\Omega)=\hat H^{(m)}_{ \mathit{s}}(\Omega)+\hat H^{(m)}_{ \mathit{a}}(\Omega)$, given by:
\begin{eqnarray}
\hat H^{(m)}_{ \mathit{s/a}}(\Omega)&=&\pm \alpha_{0}^{*(m-1)}\hat{a}_{1s/a}\hat{a}_{2s/a}\nonumber\\&+&(m-1)\alpha^{*(m-2)}_{0}\alpha_{1}\hat{a}^{\dagger}_{0s/a}\hat{a}_{2s/a}\nonumber \\
&+&(m-1)\alpha_{0}^{*(m-2)}\alpha_{2}\hat{a}_{0s/a}^{\dagger}\hat{a}_{1s/a}\nonumber \\
&\pm &(m-2)\alpha_{1}\alpha_{2}\hat{a}^{\dagger 2}_{0s/a}-h.c.\qquad.
\label{hamsimantissim}
\end{eqnarray}

As showed in details in the reference~\cite{MUNOZ2018}, this Hamiltonian represents a two-mode squeezing process on the  twin beams in the presence of an intense pump field, and a pair of beam splitter process between the pump and one of the generated fields in the presence of an intense mean field related with the conjugated mode.
%(and the pump in the $\chi^{(3)}$ situation).
In the case of Hamiltonian, related to the $\chi^{(3)}$ medium, one additional term is present, which is related with a squeezing process in the pump field in the presence of a pair of intense converted fields.
%%%%%%%%%%%%%%%%%%%%%%%%%%%%%%%%%%%%%%%%%%%%%%%%% OPERATOR DYNAMIC %%%%%%%%%%%%%%%%%%%%%%%%%%%%%%%%%%%%%%%

%\subsection{Operators dynamic}

%In the symmetric and antisymmetric basis,
We can use the  Heisenberg equation $d\hat{a}_{n(s/a)}(t)/dt =-i/\hbar[H_{\chi^{(m)}(s/a)},\hat{a}_{n(s/a)}^{\dagger}(t)]$ to evaluate the transformation of the field operators from the input to the output of the amplifier.
%In Section \ref{ClassAppr} we defined 
%$\kappa_{m}=2\chi^{(m)} L/ (\nn c)$ 
%$\kappa_{m}=\chi^{(m)} L/ (\nn c)$ 
%to describe the power of the pump, signal and idler beams when passing through the gain medium with length $L$. 
Calculation can be performed using an  auxiliary variable $\xi_m=\chi^{(m)} t$ giving a compact form for the Heisenberg equation
%to describe the field amplitude in the gain medium, such that $d=t(\nn c) \rightarrow \xi_m=\chi^{(m)} t$. In terms of this new variable $\xi_m$, the Heisenberg equation can be written in a compact form:
\begin{equation} \label{eq_heisemberg2}
	\frac{d \vec{\textbf{A}}_{(s/a)}}{d\xi_m} = \textbf{M}_{(s/a)} (\xi_m)\vec{\textbf{A}}_{(s/a)}, \mbox{where}
\end{equation}
%the vector $\vec{\textbf{A}}_{s/a}=\vec{\textbf{A}}_{s}\oplus \vec{\textbf{A}}_{a}$ and 
$	\vec{\textbf{A}}_{s/a} = (\hat{a}_{0(s/a)}~~ \hat{a}_{0(s/a)}^{\dagger} ~~\hat{a}_{1(s/a)} ~~\hat{a}_{1(s/a)}^{\dagger}~~ \hat{a}_{2(s/a)}~~ \hat{a}_{2(s/a)}^{\dagger} )^T$. The evolution matrix has an explicit dependence on the amplitude of the fields inside the crystal, that evolve along the propagation, as evaluated in the previous section, %$\textbf{M}_{(s/a)} (\xi_{m})$:

\begin{widetext}
 \begin{footnotesize}
 \setlength{\arraycolsep}{0.1pt}
\renewcommand{\arraystretch}{0.1}
 \begin{eqnarray}
 \textbf{M}_{(s/a)}&=&\\&\xi_{m}&\left(
\begin{array}{cccccc}
 0 & \mp 2(m-2)\alpha_{\omega 1}\alpha_{\omega 2} &  -(m-1)\alpha^{*(m-2)}_{\omega 0}\alpha_{\omega 2}  & 0 & - (m-1)\alpha^{*(m-2)}_{\omega 0}\alpha_{\omega 1}  & 0  \\
 \mp 2(m-2)\alpha_{\omega 0}\alpha^{*}_{\omega 2} & 0 & 0  & -(m-1)\alpha^{(m-2)}_{\omega 0}\alpha^{*}_{\omega 2}  & 0  & - (m-1)\alpha^{*(m-2)}_{\omega 0}\alpha^{*}_{\omega 1}  \\
(m-1)\alpha^{(m-2)}_{\omega 0}\alpha^{*}_{\omega 2}  & 0 & 0  & 0  & 0  & \pm \alpha^{m-1}_{\omega 0}\\
0  & (m-1)\alpha^{*(m-2)}_{\omega 0}\alpha_{\omega 2}  & 0 & 0  & \pm \alpha^{m-1}_{\omega 0}  & 0  \\
 (m-1)\alpha^{(m-2)}_{\omega 0}\alpha^{*}_{\omega 1}  &0 &0 & \pm \alpha^{*(m-1)}_{\omega 0}  & 0  & 0 \\
 0  & (m-1)\alpha^{*(m-2)}_{\omega 0}\alpha_{\omega 1} & \pm \alpha^{*(m-1)}_{\omega 0}  & 0  & 0 & 0  \nonumber \\
\end{array}
 \right). 
\end{eqnarray}
\end{footnotesize}
\end{widetext}
Solving the differential  Eq. (\ref{eq_heisemberg2}), the result can be written as:
\begin{equation}\label{gain}
\vec{\textbf{A}}_{(s/a)}\mid_{\xi_m=\kappa^{(m)}} = \textbf{G}_{m(s/a)} \vec{\textbf{A}}_{(s/a)}\mid_{\xi_m=0},
\end{equation}
with 
%where all beams conversion rate are given by \autoref{difpower} and integrating  Eq. \ref{eq_heisemberg2} yields:
\begin{equation}
	\textbf{G}_{m(s/a)}= exp\left(\int_0^{\kappa^{(m)}} d\xi_m\textbf{M}_{m(s/a)}(\xi_m)\right).
 \label{eq_ganho}
\end{equation}
%Note that, the pump power decreases quickly through the gain medium for second or third non-linearity.

Most of the works until now considered that the evolution of the field inside the crystal is negligible due to the low total power variation of signal of this systems~\cite{Debuisschert:93}, a valid situation for a small gain, typical situation found in closed cavities. The integration of~ Eq. (\ref{eq_ganho}) allows us to study the behavior of the fields inside two different gain media without this consideration enabling the accurate study in the open cavity regime \cite{alvaro2020}.
%%%%%%%%%%%%%%%%%%%%%%%%%%%%%%%%%%%%%%% Optical Cavity %%%%%%%%%%%%%%%%%%%%%%%%%%%%%%%%%%%%%%%%%%%%%%

%%% Figura Cavidade %%%
\begin{figure}[b]
	\centering
	\includegraphics[scale=1.3]{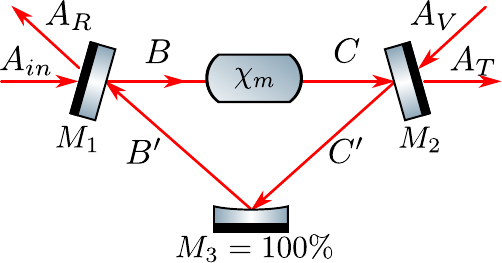}
	%{Cavitywithgainmedium_operators.pdf}
\caption{\label{cavfp} Representation of the fields inside the cavity with a gain medium $\chi^{(m)}$. $M_1$ and  $M_2$ are the input and output mirrors, with reflectivities (transmissivities) coefficients $\mathbf{R}$ ($\mathbf{T}$) and $\mathbf{R}^{'}$($\mathbf{T}^{'}$).}
\end{figure}

In order to evaluate the behavior of the field fluctuations in a round trip inside the cavity (Fig. \ref{cavfp}) the procedure described in~\cite{MUNOZ2018} was adapted to a format that can be used considering both $\chi^{(2)}$ and $\chi^{(3)}$ as the gain medium inside the cavity (see Appendix \ref{transform} for further details).
The output field $\bf{\hat{A}_{R}}$ from the cavity is directly related to the incident field $\bf{\hat{A}_{in}}$ and the additional vacuum field $\bf{\hat{A}_v}$, associated to spurious losses of the cavity, by the relation
%The procedure relates the field vector reflected by the cavity, $\bf{\hat{A}_{R}}$, the vector of the transmitted field by the output mirror, $\bf{\hat{A}_T}$, and the vacuum vector, $\bf{\hat{A}_v}$, associated to losses of the cavity, by the relation:
\begin{equation}\label{cavityBS}
   \vec{\textbf{A}}_R = \textbf{R}_{\kappa}\vec{\textbf{A}}_{in} + \textbf{T}_{\kappa}\vec{\textbf{A}}_{\nu}.
\end{equation}
where $\textbf{R}_{\kappa}$ and $\textbf{T}_{\kappa}$ are the effective reflection and transmission matrices of the OPO cavity, accounting for all the mirror coupling and the gain transformation described by Eq. (\ref{eq_ganho}).

Knowing the OPO output fields, the correlation between the output quadratures can be analyzed. Following the analysis in the reference~\cite{MUNOZ2018} we performed a complete description of the covariance matrix of the Hermitian operators $\hat{p}_{\omega_{n}}$ and $\hat{q}_{\omega_{n}}$ of pump, signal and idler modes, that satisfies the commutation relation $[\hat{p}_{\omega}, \hat{q}_{\omega^{'}}^{'}]= 2 i \delta (\omega-\omega^{'})$  and are related with the operators $\hat{a}_{\omega}$ and $\hat{a}_{\omega}^\dagger$ by $\hat{p}_{\omega}= (\hat{a}_{\omega}+\hat{a}_{\omega}^{\dagger})/2$ and $\hat{q}_{\omega}= i(\hat{a}_{\omega}-\hat{a}_{\omega}^{\dagger})/2$. The covariance matrix of the reflected field is given by:
\begin{equation}\label{eq:matCovar_main}
\textbf{V}_R =\tilde{ \textbf{R}}_\kappa \textbf{V}_{in}\tilde{\textbf{R}}_\kappa^{-1}+\tilde{\textbf{T}}_\kappa \textbf{V}_{\nu}\tilde{\textbf{T}}_\kappa^{-1},
\end{equation}
where $\textbf{V}_{in}$ is the input covariance matrix of the pump, signal and idler fields, and $\textbf{V}_{\nu}$ is the covariance matrix related with the input vacuum modes. We considered that the input covariance matrix represents a coherent state, $\textbf{V}_{\nu}=\textbf{V}_{in}=\textbf{I}$. Details of the calculation procedure can be found in  \cite{MUNOZ2018}.

\section{Analysis of the covariance matrix\label{secent}}

The covariance matrix $\textbf{V}_R$ gives a unique description of the system, and for a Gaussian state is equivalent to the determination of the density operator \cite{Simon}. It gives all the possible information about the system, including squeezing and entanglement of the fields. In what follow, we will make a detailed analysis of the quantum features that can be found in the different configurations, DROPO and TROPO, $\chi^{(2)}$ and $\chi^{(3)}$.

In order to give a general view, we have chosen a reflectance of $R=85\%$ for the output coupler, and no additional loss in the cavity. Analysis frequency is chosen to be half of the cavity bandwidth $BW$: $\Omega=0.5 (1-R)/\tau$, where $\tau$ is the round trip time of the wave inside the cavity. TROPO is chosen to be with all the reflectances identical.

%%%%%%%%%%%%%%%%%%%%%%%%%%%% Analise grafica %%%%%%%%%%%%%%%%%%%%%%%%%%%%%%%%%%%%%%%%%%%%%%%%%%%%%%%%%%
\subsection{Source of squeezed states}
%%%%%%%%%%%%%%%%%%%%%%%%%%%% DROPO_Chi2_Chi3_VARIANCES%%%%%%%%%%%%%%%%%%%%%%%%%%%%%%%%%%%%%%%%%%%%%%%%%%%%%%%%%%

If we want to observe the noise compression, as evaluated for instance in \cite{Fabre1997}, we may restrict the study to the covariance of the symmetric variables \cite{PhysRevA.88.052113}. From the symmetry between signal and idler modes, only the presentation of the variances of one of these fields is necessary.
The quadratures associates to the amplitude and phase are represented by $\Delta p_{si}$ and $\Delta q_{si}$, with $i=\{0,1,2\}$ for pump, signal and idler fields. 

\begin{figure}[b]
	\centering
	\includegraphics[scale=0.30]{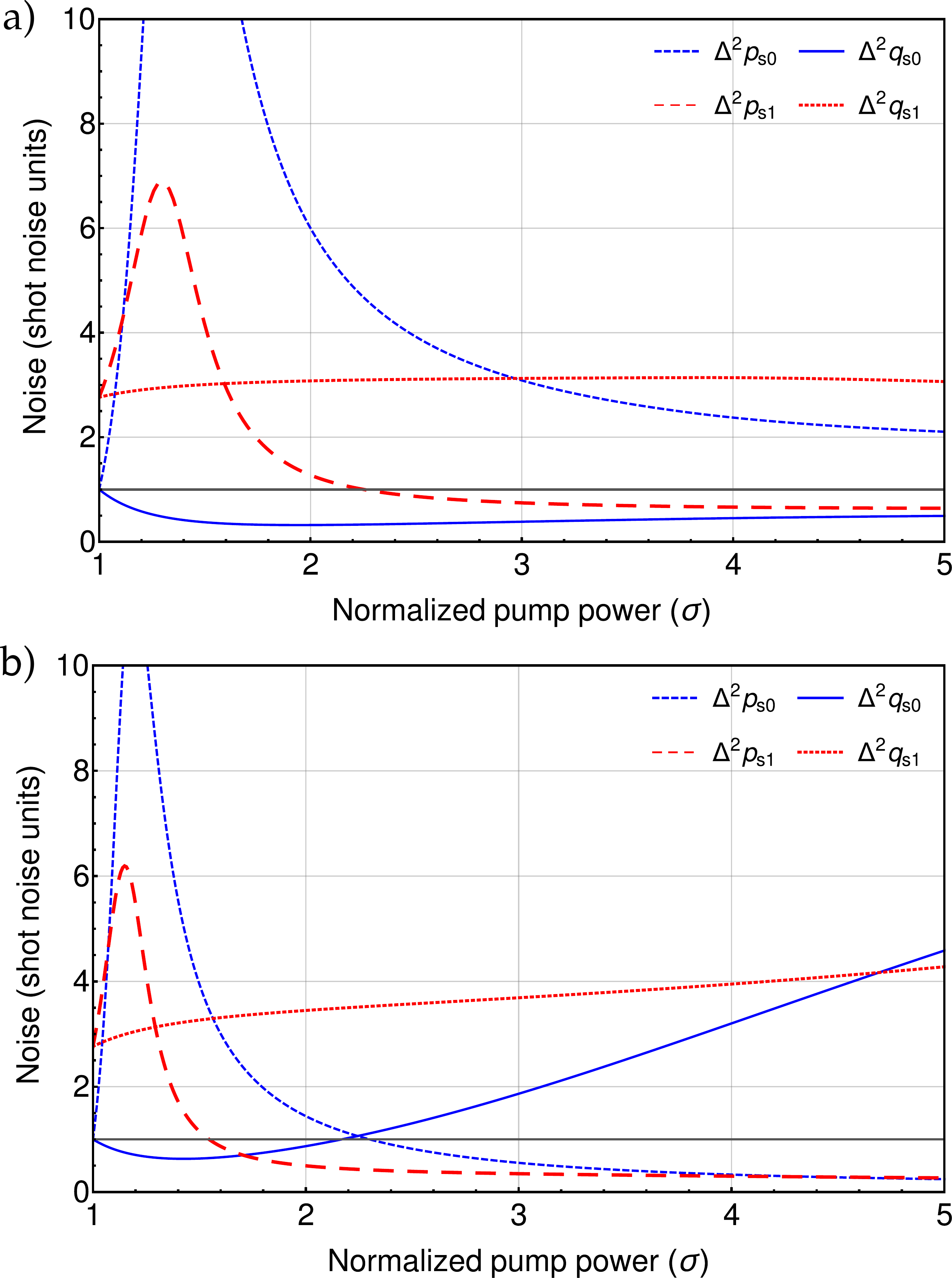}
\caption{\label{VarSimchi23} DROPO: Diagonal terms of $V_{s}$ in terms of $\sigma$ for $R_{1}=85\%$ and $\Omega= 0.5 BW$. Considering in a) a $\chi^{(2)}$ gain medium and in b) a $\chi^{(3)}$ gain medium.} 
\end{figure}
%%%%%%%%%%%%%%%%%%%%%%%%%%%%%%%%%%%%%%% TROPO %%%%%%%%%%%%%%%%%%%%%%%%%%%%%%%%%%%%%%%%%%%%%%%%%%

%%%%%%%%%%%%%%%%%%%%%%%%%%% TROPO Chi2_Chi3_VARIANCES_figure %%%%%%%%%%%%%%%%%%%%%%%%%%%%%%%%%%%%
 \begin{figure}[hb]
	\centering
	\includegraphics[scale=0.30]{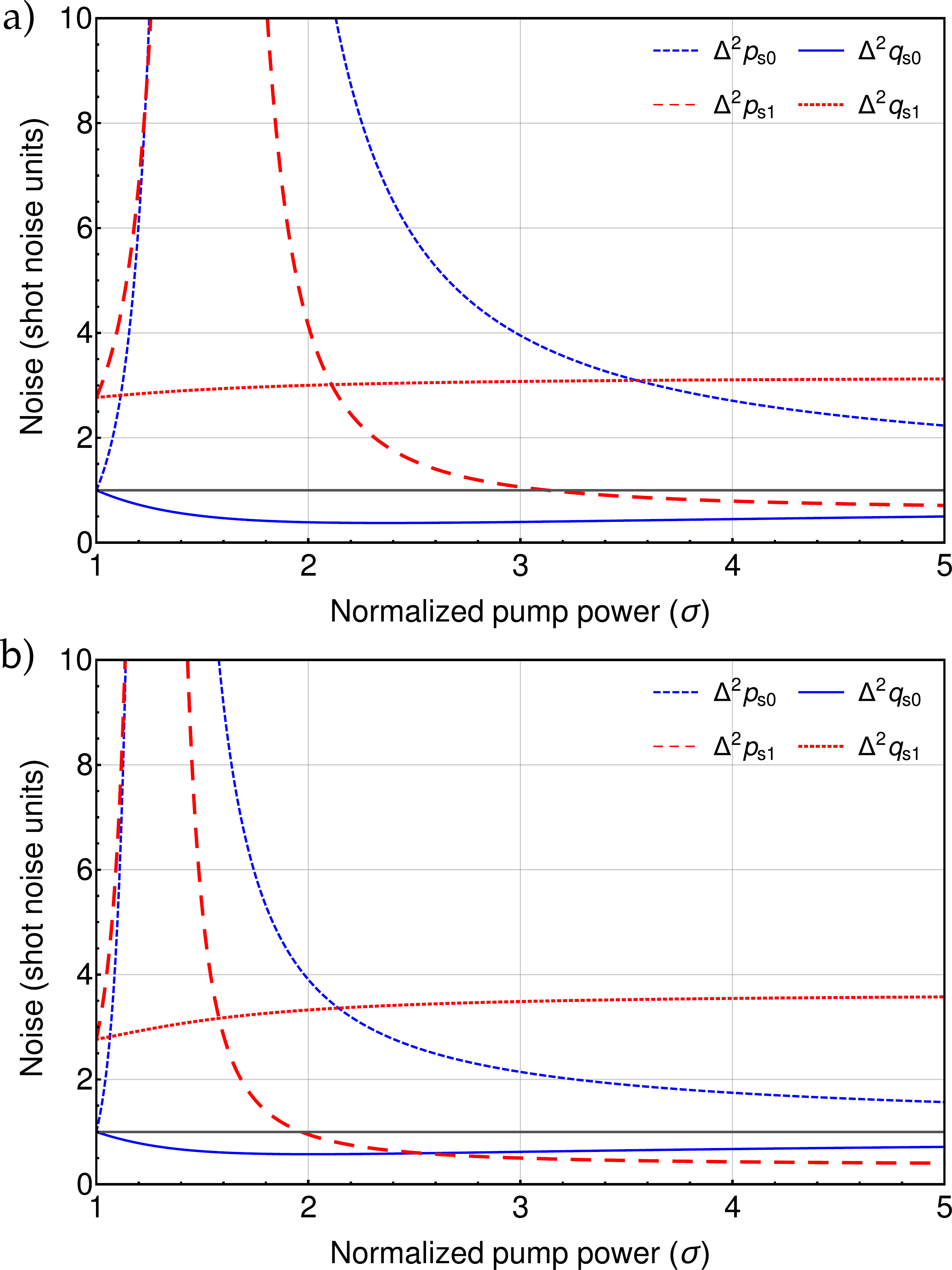}
 \caption{\label{fig:TROPOVarSim} TROPO: Diagonal terms of $V_{s}$ in terms of $\sigma$ for $R_{1}=85\%$ and $\Omega= 0.5 BW$. Considering in a) a $\chi^{(2)}$ gain medium and in b) a $\chi^{(3)}$ gain medium.}
 \end{figure}

Individual variances for the DROPO are presented in~Fig. \ref{VarSimchi23}, for  $\chi^{(2)}$ and $\chi^{(3)}$ gain medium.
Equivalent results for the TROPO are presented in Fig.~\ref{fig:TROPOVarSim}.
There are many common features on these curves. The compression of the phase noise of the pump, $\Delta^2 q_{s0}$, as observed in \cite{Fabre1997}, is verified not only in Fig.~\ref{fig:TROPOVarSim}a, but in all the distinct configurations. It is interesting to notice that while the compression is limited to 0.5 in the $\chi^{(2)}$ TROPO \cite{Fabre_1990}, the $\chi^{(2)}$ DROPO can beat this value. On the other hand, while compression of this quadrature in presented in $\chi^{(3)}$ OPO, they are not so effective as squeezers for the pump. In fact, for the $\chi^{(3)}$ DROPO, noise compression is even limited to the range of $\sigma < 2.2$. But then we have a curious feature: for $\sigma < 2.3$, the pump amplitude $\Delta^2p_{s0}$ becomes squeezed. This difference in behavior between $\chi^{(2)}$ and $\chi^{(3)}$ OPOs can be explained by the additional term in Eq.~(\ref{hamsimantissim}), giving the compression operator acting on the pump, associated to the mean converted fields inside the cavity. This effect should compete with the usual dynamics of the phase noise compression provided by the back conversion of the signal and idler fields into the pump mode described in \cite{Fabre1997}. For a strong field, it should beat the phase compression.

As for the converted fields, they present a nearly perfect thermal state right above the threshold, but for an increasing pump power, while phase noise $\Delta^2q_{s1}$ grows smoothly, the amplitude noise $\Delta^2p_{s1}$ presents a strong peak, that is much more pronounced for the TROPO, and almost coinciding with the peak noise for the pump amplitude $\Delta^2p_{s0}$. On the other hand, above a certain value, the noise drops and we eventually have noise compression for this field. While this effect was already predicted in the literature for the $\chi^{(2)}$ TROPO, to the best of our knowledge is was not observed yet. A good reason could be the fact that is should appear above $\sigma=3$, situation where the thermal effects will become dramatic in optical crystals, and the intense fields that are produced will elude the usual homodyne techniques for noise measurement. It would be necessary to use self-homodyning, as done in \cite{Villar2005}, for its observation. Nevertheless, the use of $\chi^{(3)}$ gain medium on a DROPO reduces the value of the necessary pump power for reaching the squeezed output. In fact, for a $\chi^{(2)}$ DROPO we reach a significant noise reduction to the shot noise level, that was even observed in preliminary studies using an OPO with atomic vapor as the gain medium \cite{alvaro2020}, although squeezing was not yet verified. The strong compression for the  amplitudes of pump, signal and idler field is a dramatic demonstration of the role of pump depletion even in a single pass of the beam through the crystal. 

\begin{figure}[b]
\centering
\includegraphics[scale=0.30]{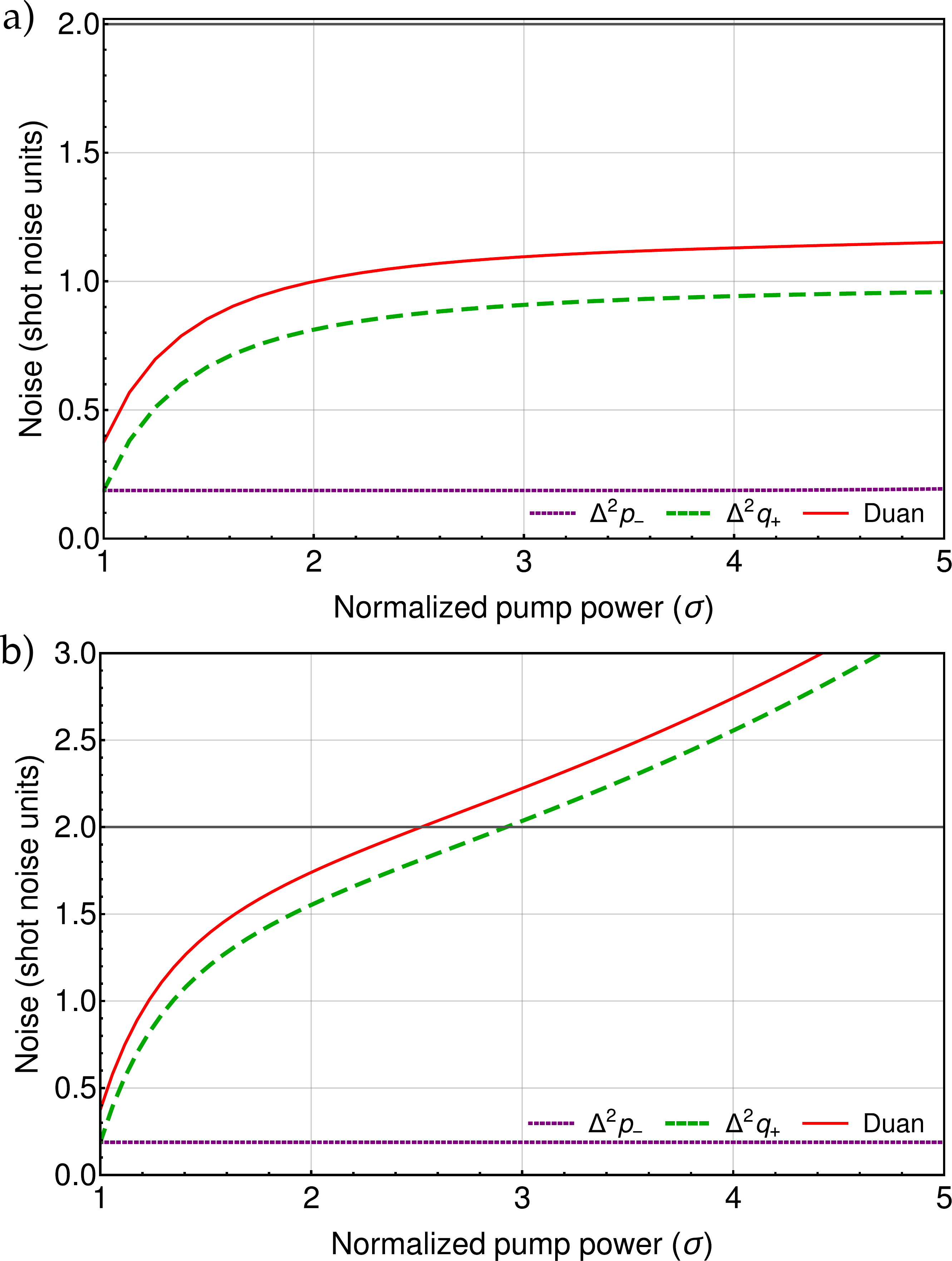}
\caption{\label{Duanchi23} DROPO: EPR inequality (continuum, orange), $\Delta^{2} p_{-}$ (dashed,purple) and $\Delta^{2} q_{+}$ (dashed, green) in terms of $\sigma$ for $R_{1}=85\%$ and $\Omega=0.5 BW$. Considering in a) a $\chi^{(2)}$ gain medium and in b) a $\chi^{(3)}$ gain medium. The grey solid line represents the limit value of the inequality given by~ Eq. (\ref{eq_duan}).}
\end{figure} 

\begin{figure}[hb]
\begin{center}
\includegraphics[scale=0.30]{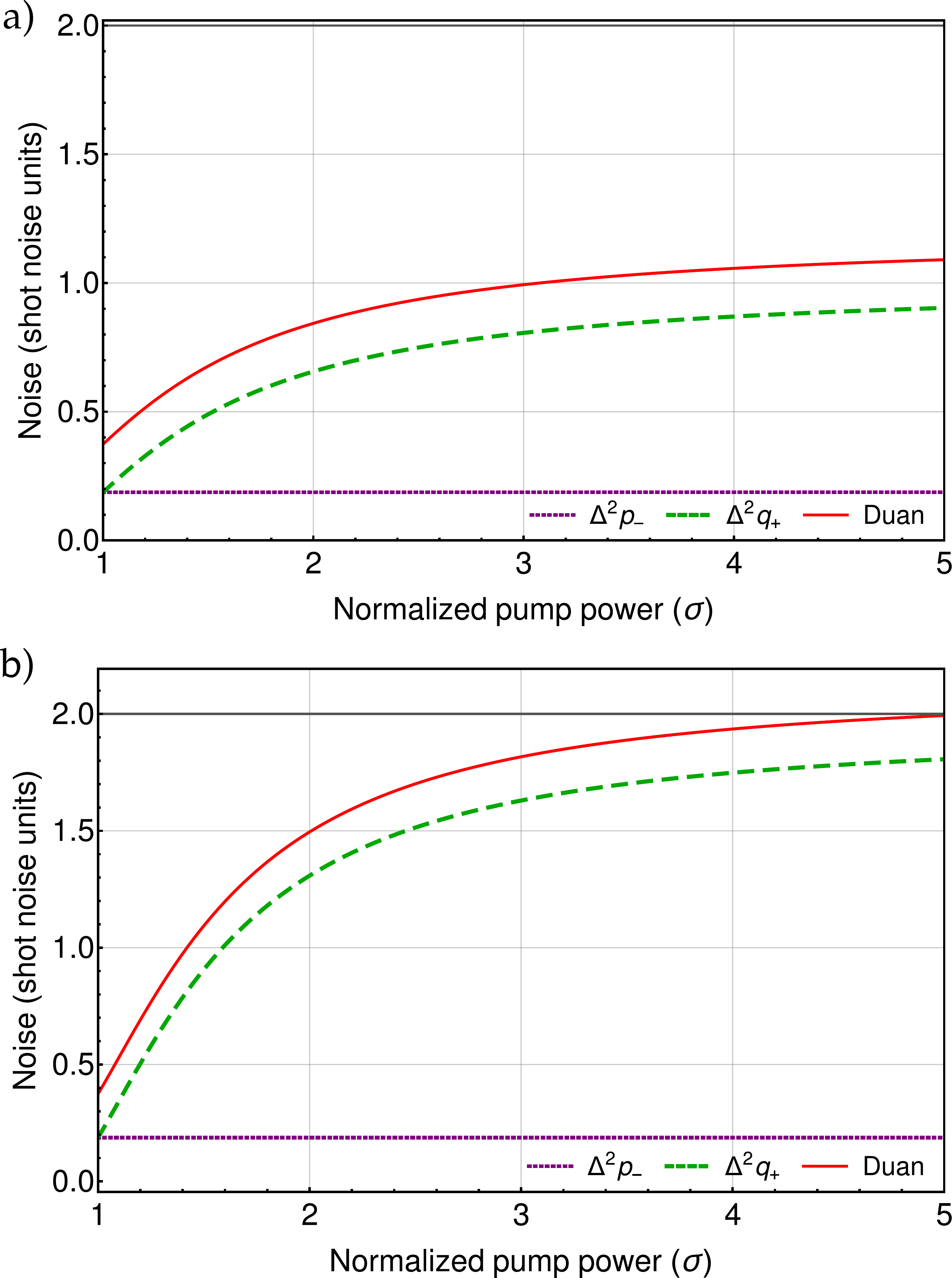}
\caption{TROPO: EPR inequality (continuum, orange), $\Delta^{2} p_{-}$ (dashed,purple) and $\Delta^{2} q_{+}$ (dasehd, green) in terms of $\sigma$ for $R_{1}=85\%$ and $\Omega=0.5 BW$. Considering in a) a $\chi^{(2)}$ gain medium and in b) a $\chi^{(3)}$ gain medium. The grey solid line represents the limit value of inequality~ Eq. (\ref{eq_duan}).}
\label{fig:TROPODuan}
\end{center}
\end{figure}
%%%%%%%%%%%%%%%%%%%%%%%%%%%%%%%%%%%%%%%%%%%% FIGURE DUAN TROPO %%%%%%%%%%%%%%%%%%%%%%%%%%%%%% 

%%%%%%%%%%%%%%%%%%%%%%%%%%%%%%%%%%%%%% QUANTUM CORRELATION CRITERIA %%%%%%%%%%%%%%%%%%%%%%%%%%%%%%%%%%%%%%
%basis (singular) bases(plural) 
\subsection{Bipartite Entanglement}

Two mode entanglement, the basic resource for quantum information processing, can be directly observed from the second order momenta \cite{Reid1989,Duan2000}. 
%criterion (singular) criteria (plural)
%%%%%%%%%%%%%%%%%%%%%%%%%%%%%%%%%%%%%%%%%%%%%% DUAN %%%%%%%%%%%%%%%%%%%%%%%%%%%%%%%%%%%%%%%%%%%%%%
%\subsection{DGCZ - EPR like states} 
In fact, noise compression in the Einstein-Podoslky-Rosen type operators criteria~\cite{Duan2000} is a sufficient condition for a successful teleportation of a quantum state between two sites \cite{furusawa1998}. This DGCZ criterion can be expressed as an inequality of the form
\begin{equation}
\Delta^{2}p_{-}+\Delta^{2} q_{+}>2,
\label{eq_duan}
\end{equation}
where the EPR-type variables are $\Delta^{2} p_{-}=(p_{s1}- p_{s2})/\sqrt{2}$ and $\Delta^{2} q_{+}=(q_{s1}+ q_{s2})/\sqrt{2}$. If the variances of these lines combinations of quadratures violates the inequality, the bipartition $\{1,2\}$ is necessarily entangled.
%%%%%%%%%%%%%%%%%%%%%%%%%%%%%%%%%%%%%%%% FIGURE DUAN  DROPO %%%%%%%%%%%%%%%%%%%%%%%%%%%%%%%%%%%%%%%%%

Twin beams produced by OPOs are a regular source of entangled states \cite{Peng1992,Villar2005}. In what follows, we will evaluate the noise compression of the correlated intensities and the anti-correlated phases of the fields generated in distinct cavity configurations.
Fig.~\ref{Duanchi23} presents the behavior of~ Eq. (\ref{eq_duan}) considering the DROPO, while Fig.~\ref{fig:TROPODuan} present the results for the TROPO.
An outstanding result is the robustness of the twin beam correlation \cite{Heidmann1987}. The subtraction of the amplitudes for all the four configurations is the same, and directly related only to the cavity bandwidth and detection efficiency - the squeezing level depends on the fraction of the twin photons, that are generated by the parametric conversion, that is detected. Therefore, this variance is independent of the pump power, even though the variance of each field may change dramatically, from excess noise to squeezing, as seen in the previous section.

On the other hand, phase anti-correlation, associated to the noise compression in $\Delta^2 q_+$, is more fragile, and depends strongly on the pump power \cite{Villar2005}. Starting from the same level as the $\Delta^2 p_-$ close to the threshold, it had a monotonic increase. Here the effect of the gain medium is very relevant: while $\chi^{(2)}$ OPOs have a limit where this variance asymptotically reaches the vacuum level for increasing pump power ($\Delta^2 q_+<1$), $\chi^{(3)}$ OPOs will cross this limit at very low pump power, $\sigma\simeq1.5$. Moreover, the loss of noise compression for growing pump power is more pronounced in the DROPOs, when compared to TROPOs, for both gain medium, and should be considered on the development of entangled bipartite sources.

As a result, DGCZ inequality is violate for all the value range for the $\chi^{(2)}$ OPOs, but it is satisfied only up to a certain level of pump power for $\chi^{(3)}$ OPOs. This is a main limitation of this system for bipartite entanglement, but since we are dealing here with pure states, the loss of entanglement in this two mode partition can be understood as their coupling to other modes of the system, as we should see in the next subsection. 

\begin{figure}[ht]
	\centering
	\includegraphics[scale=0.30]{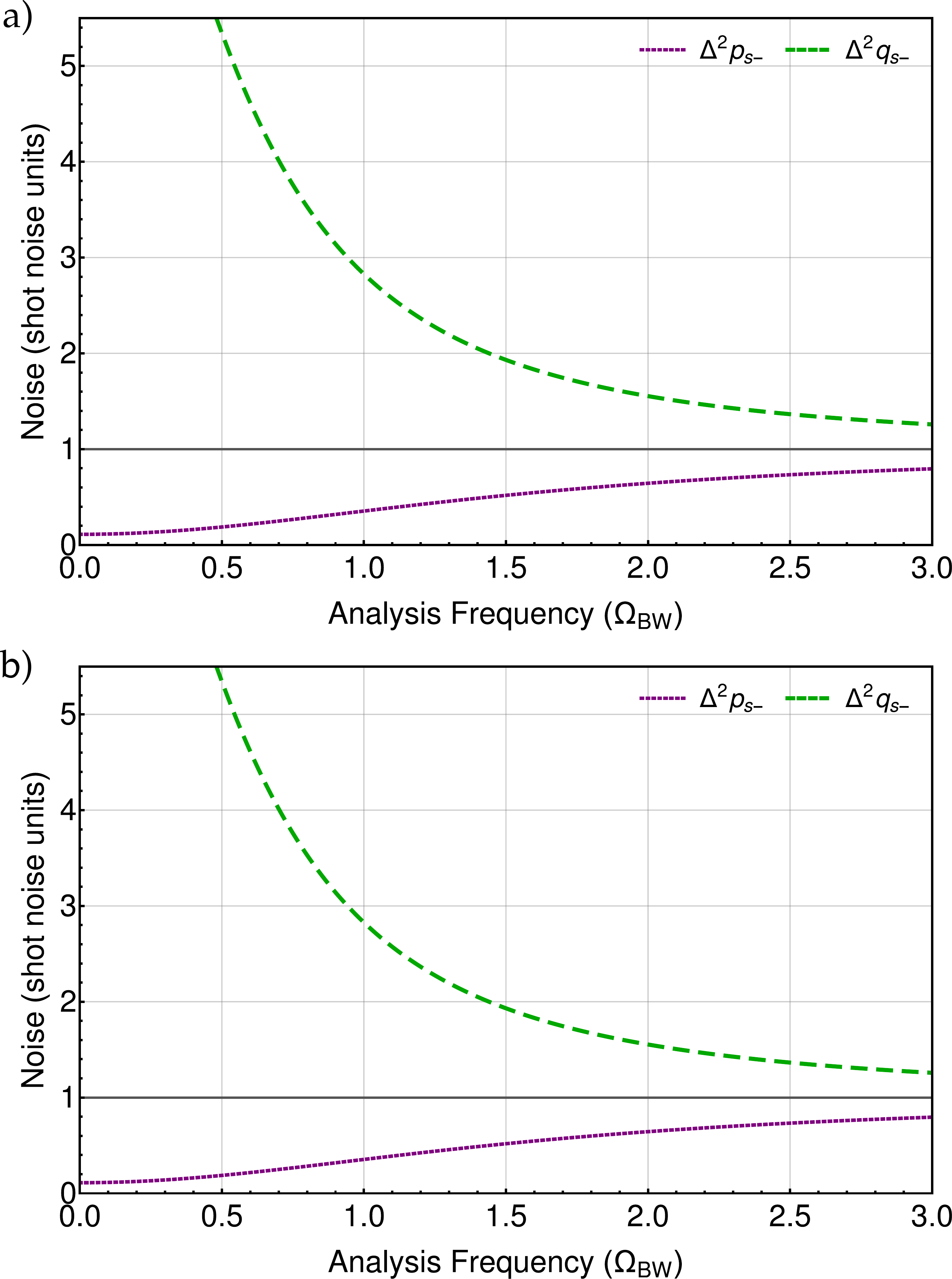}
 \caption{\label{fig:DROPOminusomega} DROPO: $\Delta^{2}p_{s-}$ and $\Delta^{2}p_{q-}$ in terms of $\Omega$ (normalized by the cavity bandwidth $BW$) for $R_{1}=85\%$ and $\Omega_{BW}=0.5$. Considering in a) a $\chi^{(2)}$ gain medium and in b) a $\chi^{(3)}$ gain medium.}
 \end{figure}

Another important analysis is related with the behavior of the noise spectrum in terms of the analysis frequency normalized by the cavity bandwidth~$\Omega_{BW}$. As an example, we focus on the substraction of the signal and idler fields. It is very well known that in $\chi^{(2)}$ OPOs the behavior of the phase ($\Delta^{2}q_{s-}$) and amplitude ($\Delta^{2}p_{s-}$) are insensitive to the pump power, and depends only on the analysis frequency. Noise compression in  $\Delta^{2}p_{s-}$ will follow a Lorenztian, of width given by the cavity bandwidth, and for a lossless system, we have $\Delta^{2}p_{s-}\Delta^{2}q_{s-}=1$.
The same situation is verified for the DROPO, as can be observed in~Fig. \ref{fig:DROPOminusomega}. This response was recently been observed for a $\chi^{(3)}$ DROPO with atomic vapor \cite{alvaro2020}, where the twin beam generation has showed to be independent of the pump power, and follow a Lorentzian shape.

%We have presented here simulations regarding the Duan criterion~ Eq. \ref{eq_duan} for both DROPO and TROPO and in each case considering the $\chi^{(2)}$ and $\chi^{(3)}$ gain media. In all the cases, the systems present entanglement between the signal and idler beams for at least $\sigma~2.5$, showing the importance of OPOs as a source of entanglement states of the light.
%%%%%%%%%%%%%%%%%%%%%%%%%%%%%%%%%%%%%%%  PPT_THEORY %%%%%%%%%%%%%%%%%%%%%%%%%%%%%%%%%%%%%%%%%%%%%%%%%%%%%%%%%%%%%%
\subsection{Tripartite Entanglement}

While DGCZ criterion is a useful test for bipartite entanglement, it is not both necessary and sufficient on its usual form (Eq. \ref{eq_duan}). On the other hand, positivity under partial transposition (PPT) was showed to be a necessary and sufficient criterion not only for bipartite Gaussian states~\cite{Simon}, but for $1\times N$ bipartitions as well \cite{Werner2001}.

Partial transposition operation in a CV system is like a mirror reflection in the phase space, acting only in a partition. When a transposition operator is applied on the density operator, the corresponding Wigner function transforms as $W( {\bf{x}})\rightarrow  W(\mathbf{\Gamma} {\bf{x}})$, with $\mathbf{{x}}=({p}_{1},{q}_{1},{p}_{2},{q}_{2})$, we have the substitution of $\mathbf{{x}}\rightarrow\mathbf{\Gamma} {\bf{x}}$, with $\mathbf{\Gamma}=diag(1,1,1,-1)$. 
If the new Wigner function does not correspond to physical density operator, the system is entangled. On the other hand, if it does and the Wigner function is Gaussian, then we know that the bipartition is separable, PPT can be immediately verified by the covariance matrix. 
Writing the set of commutation rules as  $[\hat{x}_{i},\hat{x}_{j}]=i\mathbf{\Omega}_{i,j}$ where $\mathbf{\Omega}=\bigoplus_{k=1}^{N} J$ and $J= \begin{pmatrix} 0 & 1 \\-1 & 0 \end{pmatrix}$, the uncertainty relation can be expressed as  $\mathbf{V}+i\mathbf{\Omega} \geq 0$. Partial transposition implies in the transformation $\mathbf{V}  \rightarrow^{PT} \mathbf{\tilde{V}}=\mathbf{\Gamma} \mathbf{V} \mathbf{\Gamma}$.
Physicality of the partially transposed covariance, 
$\mathbf{\tilde{V}}+i\mathbf{\Omega} \geq 0$. 
can be verified by the evaluation of the symplectic eigenvalues $\nu_{k}$ of $\mathbf{\tilde{V}}$ \cite{ECVS(ADESSO2007), THESISADESSO, Williamson}.
 % one can show that, if a Gaussian state is not entangled the action of the operation $\Gamma$ in the covariance matrix, 
 %$V  \rightarrow^{PT} \tilde{V}=\Gamma V \Gamma$ the uncertainty principle is valid,
 %If this condition is not obeyed, the matrix $\tilde{V}$ does not represent a physical state, meaning that the Gaussian state represented by the covariance matrix $V$ is entangled. The analysis can be simplify using the symplectic eigenvalues $\nu_{k}$ of $\tilde{V}$ \cite{ECVS(ADESSO2007), THESISADESSO, Williamson}. 
\begin{equation}
\nu_{k}=\sqrt{\left[Eigenvalues(\mathcal{V})\right]_k}\,\, ,
\label{eq_PPT}
\end{equation}
where $\mathcal{V}=-(\mathbf{\tilde{V}\Omega})^{2}$. When $\nu_{k}\geq1~\forall~k$ the transformed matrix $\mathbf{\tilde{V}}$ is physical.
It follows from this conditions that if the minimum symplectic eigenvalue $\nu<1$, the covariance matrix $\mathbf{{V}}$ corresponds to an entangled state, and the minimum symplectic eigenvalue is an entanglement witness for the given bipartition.
The treatment can be extended to multipartite states, and for a $1\times N$ bipartition, this is a necessary and sufficient condition to demonstrate entanglement for Gaussian states \cite{Werner2001}.

%%%%%%%%%%%%%%%%%%%%%%%%%%%%%%%%%%% DROPO PPT %%%%%%%%%%%%%%%%%%%%%%%%%%%%%%%%%%%%%%%%%%%%%%%%
\begin{figure}[t]
	\centering
	\includegraphics[scale=0.30]{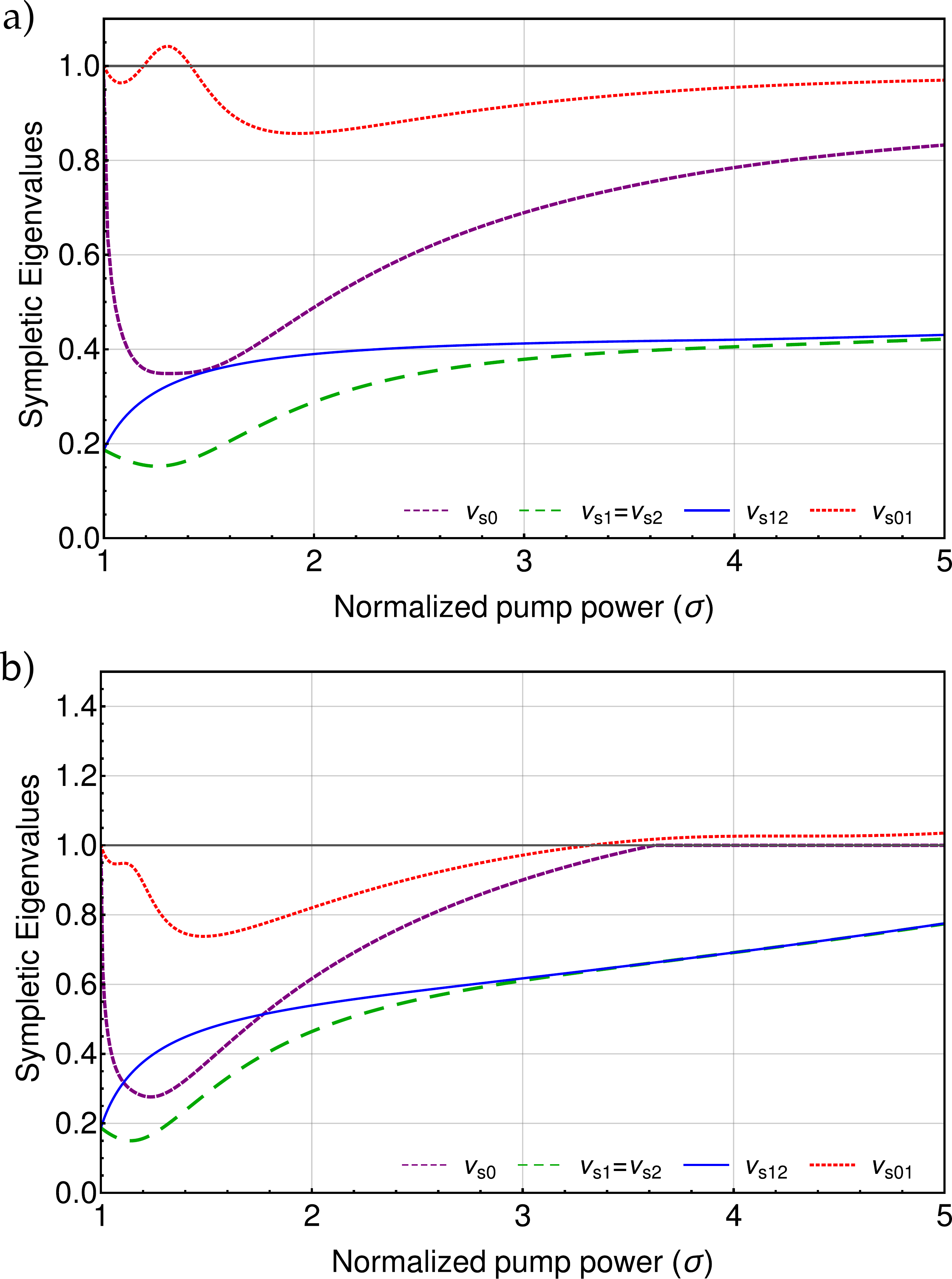}
\caption{\label{pptchi23} DROPO: Minimum symplectic eigenvalues $\nu_{k}$ obtained from $\textbf{V}_{s}$ in terms of $\sigma$ for $R_{1}=85\%$ and $\Omega= 0.5 BW$. Considering in a) a $\chi^{(2)}$ gain medium and in b) a $\chi^{(3)}$ gain medium. The grey solid line represents the limit value of Eq.~ (\ref{eq_PPT}).}
\end{figure}

\begin{figure}[ht]
	\begin{center}
		\includegraphics[scale=0.30]{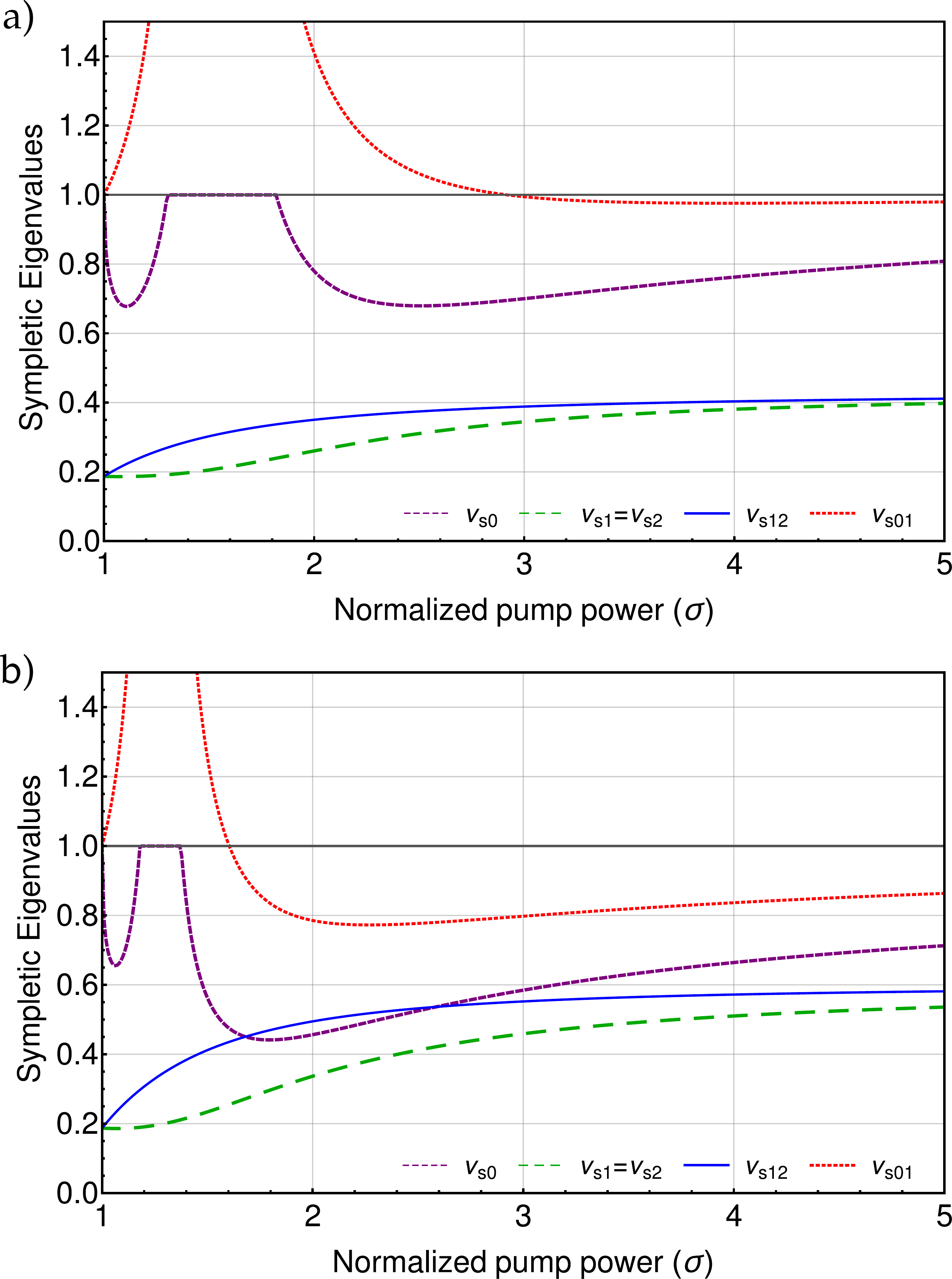}
		\caption{TROPO: Minimum symplectic eigenvalues $\nu_{k}$ obtained from $\textbf{V}_{s}$ in terms of $\sigma$ for $R_{1}=85\%$ and $\Omega= 0.5 {BW}$. Considering in a) a $\chi^{(2)}$ gain medium and in b) a $\chi^{(3)}$ gain medium. The grey solid line represents the limit value of Eq.~(\ref{eq_PPT}).}
		\label{fig:TROPOPPT}
	\end{center}
\end{figure}

For this subsection, we will keep our analysis on the subspace of the symmetric covariance matrix, as it will reflect the kind of entanglement usually observed when each beam issued from the OPO is treated as a single mode \cite{Coelho823}. For three modes, we have three possible bipartitions $1\times2$ and the tripartite entanglement is verified if  $\nu<1$ for all of them. Since the Simon-PPT criteria is necessary and sufficient, we will apply it as well to the subsystems formed by pairs of the beams, comparing the conclusions to those inferred from the DGCZ criterion.

In~Figs.~\ref{pptchi23}  and~ \ref{fig:TROPOPPT} the behavior of the symplectic eigenvalues for the DROPO  and TROPO are presented. For three modes, we have the eigenvalue for the transposition of the pump, $\nu_0$, and for signal or idler, $\nu_1=\nu_2$. We plot the transposition for the possible two mode subsystems, for pump and signal (or idler) $\nu_{01}=\nu_{02}$ and the pair formed by  the converted fields $\nu_{12}$.
%, than can be directly compared to the result obtained from the DGCZ criteria presented in Figs.~\ref{Duanchi23} and \ref{fig:TROPODuan}.

It is evident that entanglement of each converted mode to the rest of the system is always verified ($\nu_1$), but the violation is reduced for increasing pump power. It is consistent with the fact that the converted fields are strongly entangled, as can be seen by $\nu_{12}$. If you compare this situation with the observed in Figs.~\ref{Duanchi23} and \ref{fig:TROPODuan}, it is clear that Eq.~(\ref{eq_duan}) fails in identifying some entangled states: as stated in \cite{Duan2000}, the condition is both necessary and sufficient only if the covariance matrix is in one of the standard form that they propose in the article. But a feature is common in both witnesses: the violation for the $\chi^{(3)}$ DROPO is rapidly reduced for a growing pump power.

%each $\nu_{k}$ was obtained by the partial transposition of the subsystem $k$ in function of $\sigma$ for $k=\{0,1,2,01,12\}$ in the symmetric basis. If $\nu_{k}$ is smaller than 1, the bipartition is entangled. Analysing the 5 possible combinations for this system using the two gain media, it is expected that all the combinations analyzed are entangled, unless of a small region between $1.2<\sigma<1.5$ for the eigenvalue $\nu_{s01}$ using the $ \chi^{(2)}$ gain medium and between $2.8<\sigma<5$ for the eigenvalues $\nu_{0}$ and $\nu_{01}$ using a $\chi^{(3)}$ gain medium. 
%%%%%%%%%%$%%%%%%%%%%%%%%%%%%%%%%%%%% TROPO PPT %%%%%%%%%%%%%%%%%%%%%%%%%%%%%%%%%%%%%%%%%$

%

Situation is more peculiar when we look at the pump mode. Entanglement of the pump with the pair of the converted fields $\nu_0$, or with just one of the fields $\nu_{01}$ is weaker than the observed for each of the down converted beams. Moreover, it can vanish in the region where we observe the peak in the amplitude fluctuations in Figs. \ref{VarSimchi23} and \ref{fig:TROPOVarSim}. This apparent loss of entanglement is not observed in other configurations of the TROPO \cite{villar2006, Coelho823}, where it remains entangled over the entire span of the pump power. The main difference in this case is the fact that all the modes have the same loss for the cavity. This loss of entanglement would be expected if we have loss of purity in the tripartite state \cite{purity}, but as we have found, the situation is more subtle, and looking in the details of the sideband modes, we have in fact hexapartite entanglement on the system \cite{barbosa2018hexapartite}.
As for the  $\chi^{(3)}$ DROPO, we can see that for sufficiently high pump power the pump is apparently disentangled from the twin pair. That is not necessarily true, if we consider now the correlation between the symmetric and the antisymmetric part of the covariance matrix.

%It is interesting to note that applying the PPT criterion considering the bipartition $i$ vs. $jk$ the symplectic eigenvalues present smaller values then in the bipartition $i$ vs $j$, Implying that has entanglement becomes more expressive when using more information about the system. For example, looking for the bipartition $\nu_{12}$  that presents entanglement in all the region analysed. Thus, when information about the pump beam is added, represented by $\nu_{1}$, the entanglement between the signal beam and the reminder system is more expressive. The same behavior is seeing in the eigenvalues $\nu_{01}$ and $\nu_{1}$, pointing here the region in which these systems are separated $\sigma \leq 1.6 $. Excluding this region, all the symplectic eigenvalues analysed are smaller than 1 implying in a tripartite entanglement between pump, signal and idler beams in the system.

% B.M. - Essa parte descrita acima é sempre verdade, logo não acho que é necessária no texto.

\subsection{Multipartite entanglement}

%%%%%%%%%%%%%%%%%%%%%% Hexapartite analysis%%%%%%%%%%%%%%%%%
So far, we have limited our analysis only to the symmetric combination of the sidebands, that was showed to be equivalent to the antisymmetric part \cite{PhysRevA.88.052113} in the OPO. As we have showed in \cite{barbosa2018hexapartite}, in the TROPO all the possible 31 bipartitions are entangled. Therefore, we will restrict the current analysis to the relevant features involving bipartitions in the symmetric/antisymmetric basis

\begin{figure}[b]
	\begin{center}
		\includegraphics[scale=0.30]{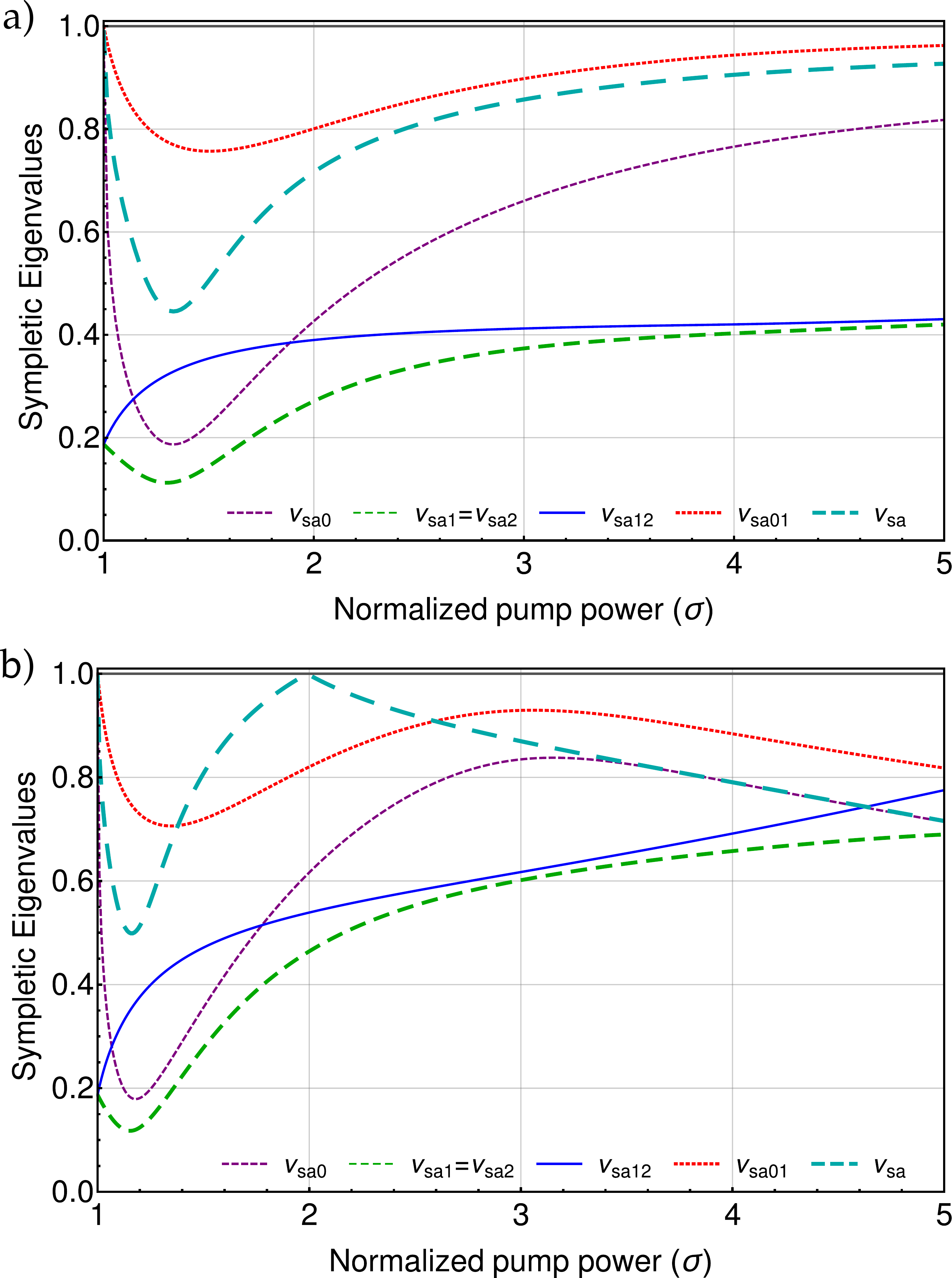}
		\caption{DROPO: Minimum symplectic eigenvalues $\nu_{sa,k}$ considering $\textbf{V}_{s/a}$ in terms of $\sigma$ for $R_{1}=85\%$ and $\Omega_{BW}=0.5$. $k = \{0, 1, 01, 12\}$ represent the subsystems of pump, signal, pump and signal, signal and idler fields respectively, and $\nu_{sa}$ is obtained by the transposition of antisymmetric pump, signal and idler fields. Considering in a) a $\chi^{(2)}$ gain medium and in b) a $\chi^{(3)}$ gain medium.}
		\label{fig:DROPOPPThexa}
	\end{center}
\end{figure}

\begin{figure}[hb]
	\begin{center}
		\includegraphics[scale=0.30]{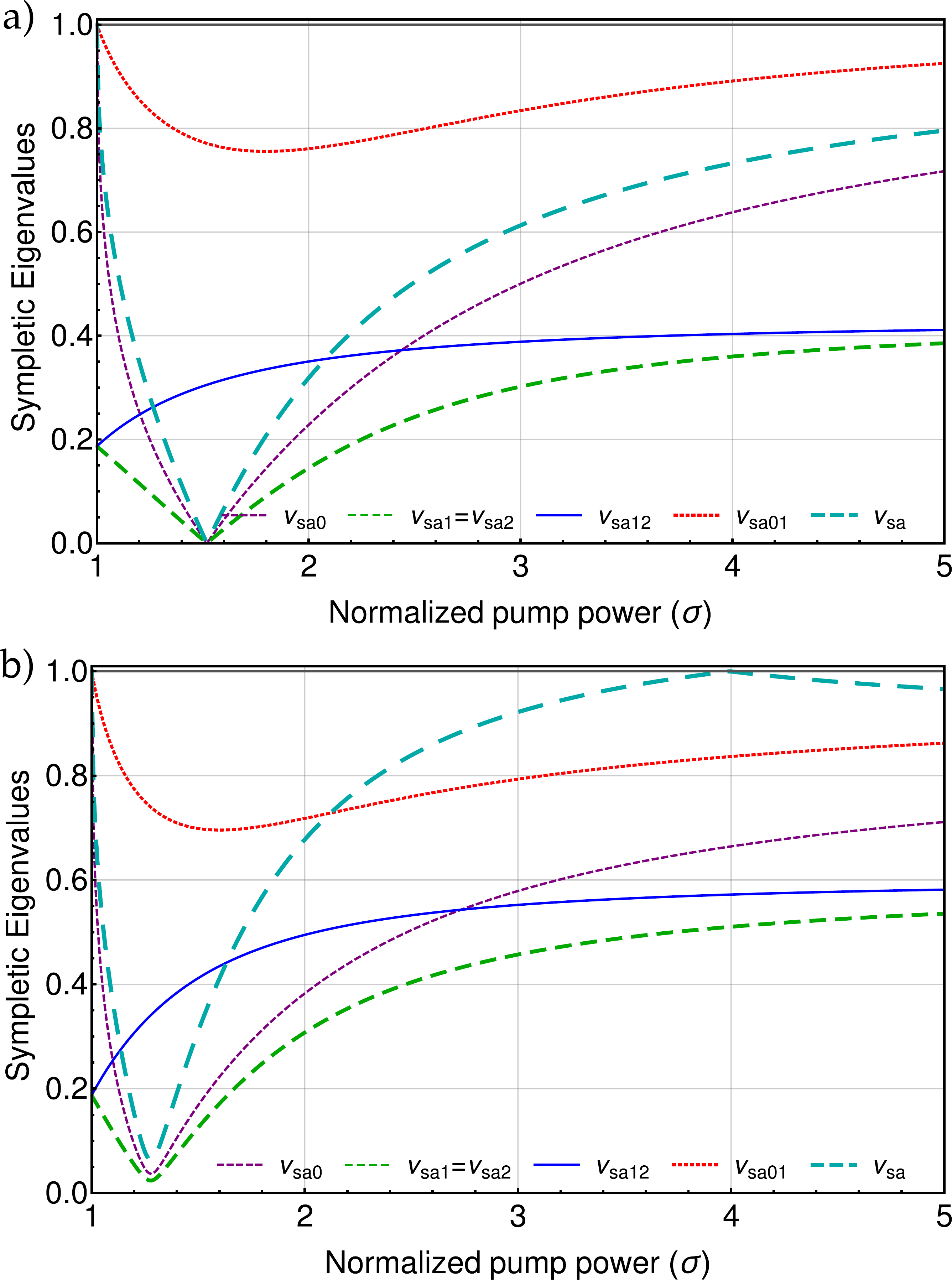}
		\caption{TROPO: Minimum symplectic eigenvalues $\nu_{sa,k}$ considering $\textbf{V}_{s/a}$ in terms of $\sigma$ for $R_{1}=85\%$ and $\Omega_{BW}=0.5$. In which $k = \{0, 1, 01, 12\}$ represent the subsystems of pump, signal, pump and signal, signal and idler fields respectively, and $\nu_{sa}$ is obtained by the transposition of antisymmetric pump, signal and idler fields. Considering in a) a $\chi^{(2)}$ gain medium and in b) a $\chi^{(3)}$ gain medium.}
		\label{fig:TROPOPPThexa}
	\end{center}
\end{figure}
%%%%%%%%%%$%%%%%%%%%%%%%%%%%%%%%%%%%%%%%%%%% DROPO vs TROPO %%%%%%%%%%%%%%%%%%%%%%%%%%%%%%%%%%%%%%%
For the six modes involved, we will explore the bipartition involving all the modes of the pump ($\nu_{s0a0}$) and all the modes of the signal or idler ($\nu_{s1a1}= \nu_{s2a2}$). It will also be relevant to explore the $\times 3$ partition, involving all the symmetric $\times$ all the antisymmetric modes ($\nu_{sa}$).

As we can see in Figs. \ref{fig:DROPOPPThexa} (for the DROPO) and \ref{fig:TROPOPPThexa} (for the TROPO), entanglement for the pump is completely recovered once we account for the complete description of the state. The same is true for signal or idler modes. The reason for this effect is clear when we consider the shared information between symmetric and antisymmetric case: entanglement is maximized, with a  particularly strong violation for $\nu_{sa}$ in the TROPO (Fig. \ref{fig:TROPOPPThexa}). We have a minimum of $\nu_{sa}$ in the region where we have a peak in the noise of the amplitudes (Figs. \ref{VarSimchi23} and \ref{fig:TROPOVarSim}). This peak is associated with a strong correlation between the symmetric and antisymmetric modes, as is showed in the Appendix \ref{App_CV_sym_antsym}. This effect is dramatically enhanced in the TROPO when the three modes have balanced losses, and is less dramatic (yet still recognizable \cite{PhysRevA.88.052113}) for the usual $\chi^{(2)}$ TROPO, where the coupling of the cavity for the pump is $\approx$ 5 times that of the downconverted modes. On the other hand, for the $\chi^{3}$ OPO, we have a unitary value for $\nu_{sa}$, at $\sigma=2$ for the DROPO and $\sigma=4$ for the TROPO. That is associated to a vanishing correlation between the partitions, as can be seen in Appendix \ref{App_CV_sym_antsym}.
%%%%%%%%%%%%%%%%%%%%%%%%%%%%%%%%%%%%%%%%%%%%%%%%%% CONCLUSION %%%%%%%%%%%%%%%%%%%%%%%%%%%%%%%%%%%%%%%%%%%%%%%%%%

Finally, instead of considering the bipartite case of pump and idler, or signal and idler, we may now consider the whole symmetric and antisymmetric combination for the pair of beams. We may compare now $\nu_{sa12}$ in Figs. \ref{fig:DROPOPPThexa} and \ref{fig:TROPOPPThexa} with $\nu_{s12}$ in
Figs. \ref{VarSimchi23} and \ref{fig:TROPOVarSim}. We cannot observe any difference, what is consistent with the absence of correlation between symmetric and antisymmetric part for signal and idler subsystems. That is not true for the pump mode, that presents relevant correlations involving the symmetric and antisymmetric part of the signal (or idler) modes as can be observed from $\nu_{sa01}$ in Figs. \ref{fig:DROPOPPThexa} and \ref{fig:TROPOPPThexa} compared to $\nu_{s01}$ in Figs. \ref{VarSimchi23} and \ref{fig:TROPOVarSim}. Entanglement between pump and signal in this case is recovered once that the correlation between symmetric and antisymmetric part are taken into account.

%Extending the analysis considering the antisymmetric bases using the PPT criterion, an hexapartite system can be analysed using the bipartitions related with the symplectic eigenvalues $\nu_{s0a0}$, $\nu_{s1a1}$, $\nu_{s01a01}$, $\nu_{s12a12}$ in the symmetric and antisymmetric bases. This analysis was done considering both DROPO and TROPO systems. The \autoref{fig:DROPOPPThexa} represents the hexapartite analysis in the DROPO configuration. We can see that for both gain media show entanglement between the subsystems.

%In TROPO configuration, all the eigenvalues presented in~\autoref{fig:TROPOPPThexa} have values smaller than 1 indicating entanglement between all subsystems, with only exception is a small region around  $\sigma=4$ where the bipartition $\nu_{sa}$ becomes separated in the $\chi^{(3)}$ TROPO.

%In this section we analysed different bipartitions possibilities between the output fields generated from an OPO considering both symmetric and antisymmetric bases. We presented a plethora of results showing the main regions in which is possible to see that the systems under study is a versatile source of entanglement states of the light. Multicolor entanglement above threshold was demonstrated for a large range of $\sigma$. The results using the $\chi^{(3)}$ gain medium open new possibilities in the study of the quantum correlations between the different beams that interact in these systems.

\section{Conclusion}

We can see clearly that the versatility of the %$\chi^{(2)}$ TR
OPO as a source of non-classical states in the continuous variable regime is not an exclusivity of the PDC process, but is also present in the case of $\chi^{(3)}$ media. Moreover, even for an extremely open cavity, reaching the limit of a single-pass of the pump through the amplifier, as is the case of the DROPO, noise compression and pump entanglement are also present.
The method presented in~\cite{MUNOZ2018} we have successfully employed reproduces these features observed in the TROPO, and put in evidence the similarities of the DROPO in comparison with the TROPO.

Our analysis here is quite distinct from the one performed for the TROPO in \cite{MUNOZ2018,barbosa2018hexapartite}, that focus on the role of individual sidebands of each one of the beams. In our current approach, we kept the analysis for six modes, but heading back to the symmetric/antisymmetric basis of these sidebands. The reason is twofold: that is the usual measurement basis, leading to the image of entanglement involving individual beams (considered as carrier plus sidebands), and gives a greater evidence of the role of the correlations between the symmetric and antisymmetric spaces. Although identical in individual information, they share a strong correlation leading to relevant entanglement. A good amount of information is lost if this correlation is ignored.

This fact is particularly evident for the dynamics of a cavity with equal losses for the pump, signal and idler, and we can in this case observe the dramatic effect of the entanglement between the symmetric and the antisymmetric modes. It makes a clear difference between the tripartite case and the hexapartite analysis. We may conclude that a detailed analysis of the sidebands is much more than just a reproduction of two equivalent tripartite systems, but rather a rich system of six strongly entangled modes.

While the parametric amplification in the single pass regime \cite{PhysRevA.78.043816} can provide strong correlations in the 4WM process, the use of a cavity can provide a great enhancement of these effects: the twin beam correlation can be as perfect as the ratio of the coupling to the overall losses of the cavity approaches unit. In this case, open cavities, as showed in \cite{alvaro2020}, are a promising source of entangled states.

The linear treatment presented here can give further guidance for the transition between the two operational regimes. It is expected that the linearization should fail close to the oscillation threshold \cite{dechoum}. If that is the case, we may go beyond the treatment described in Eq.~(\ref{hbl1}), that gives only bi-linear operators, that keep the Gaussianity of the input states. But with open cavities, thanks to the high gain, this sudden transition may be smoother and a detailed investigation of the evolution of the sate, with the measurement of higher order momenta, could be performed.

%We have studied the classical behavior of DROPO and TROPO $\chi^{(2)}$ and $\chi^{(3)}$ and comparing the output power of the fields of each system. Them, we have used this classical result as an input for the model developed in \cite{MUNOZ2018} to reconstruct the covariance matrix of the system. Through the information of the covariance matrix we study and compare the quantum fluctuations of the four different systems. We have made a complete characterization of the entanglement regions of the DROPO and TROPO $\chi^{(2)}$ and $\chi^{(3)}$. We have used Duan's test to determine regions in which the signal and idler fields can be used for application in quantum information protocols. Through the PPT test we made the characterization of the tripartite system formed by the symmetric modes of the pump, signal and idler fields and then the characterization of the entanglement regions of an hexapartite system formed by the symmetric and antisymmetric modes of the pump, signal and idler fields.

%We highlight that the Hamiltonians used in this study takes into account the depletion of the pump, an important phenomenon in the study with cavities. The $\chi^{(3)}$ scenario differs from the most recent studies regarding the four-wave mixing phenomenon, as these studies consider 4WM occurring in free space and thus, the pumping field depletion is negligible and pump is considered a classic field. It's also worth adding that the DROPO system present squeezing in the pump phase quadrature.

\section{Acknowledgments}
BAFR and RBA contributed equally for the present work.
We thank C. Gonzalez-Arciniegas and P. Nussenzveig for fruitful discussions. This work was supported by FAPESP proc. 2015/18834-0, CAPES and CNPq. B. M. is suported by FAPESP proc. 2014/27223-2.

\bibliographystyle{unsrt}
\bibliography{chi2chi2bib}

\newpage

\appendix

%%%%%%%%%%%%%%%%%%%%%%%%%%%%%%%%%%%%%TROPO -- different R's --%%%%%%%%%%%
\section{TROPO with different reflection coefficients}\label{App_TROPORdif}

%%%%%%%%%%%%%%%%%%%%%%%%%%%%%%%%%%%%%%
%\subsection{Triply Resonant Optical Parametric Oscillator $\chi^{(3)}$ with different reflection coefficients}\label{TROPO_Rdif}

An explicit procedure to evaluate the output of a $\chi^{(3)}$ TROPO,
%In order to express 
$P_{1out}$, as a function of $P_{0in}$, $R_0$ and $R_1$ can begin by expressing the total photon number, $P_{T0}$, as a function of those parameters. By taking the square of ~ Eq. (\ref{eq:sqrtP00}) and replacing $P_0(L)=P_0(0)-2\Delta P_{1}^{(3)}$ considering that $\Delta P_{1}^{(3)}=P_1(0)(1-R_1)/R_1$ in~ Eq. (\ref{eq_P_1out}), we end up with:
\begin{align}
     &4 T_0 R_0 P_{0in} \left(P_0(0)-2\frac{1-R_1}{R1}  P_1(0) \right)=\\ \nonumber
    &\left(-T_0 P_{0in}(0) + P_0(0) + 2R_0(0)\frac{1-R_1}{R1}P_1(0)\right)^2.
\end{align}
Now, we replace $P_1(0)=(P_{T0}-P_0(0))/2$ and rearranging the result relating $P_0(0), P_{T0}$ and $P_{0in}$ we have
\begin{align}\label{eq:voltacav}
     &4 R_1 T_0 R_0(0) P_{0in} \left( P_0(0)- (1-R_1)P_{T0} \right)=\\ \nonumber
    &\left((R_1-R_0) P_0(0) + R_{0} P_{T0} (1-R_{1})- R_1 P_{0in}(1-R_{0}) \right)^2.
\end{align}
Solution of~ Eq. (\ref{eq:voltacav}) for $P_0(0)$ leads to two distinct results.
%,  $P_{0sol1}(0)$ and $P_{0sol2}(0)$. 
We substitute those results in equation $P_1(0)=(P_{T0}-P_0(0))/2$ and we find two possible solutions for $P_1(0)$.
%, corresponding to it's respective $P_0(0)$, $\left[ P_{0sol1}(0),P_{1sol1}(0)\right]$, and $\left[ P_{0sol2}(0),P_{1sol2}(0)\right]$.

\begin{figure}[htbp]
	\centering
	\includegraphics[scale=0.3]{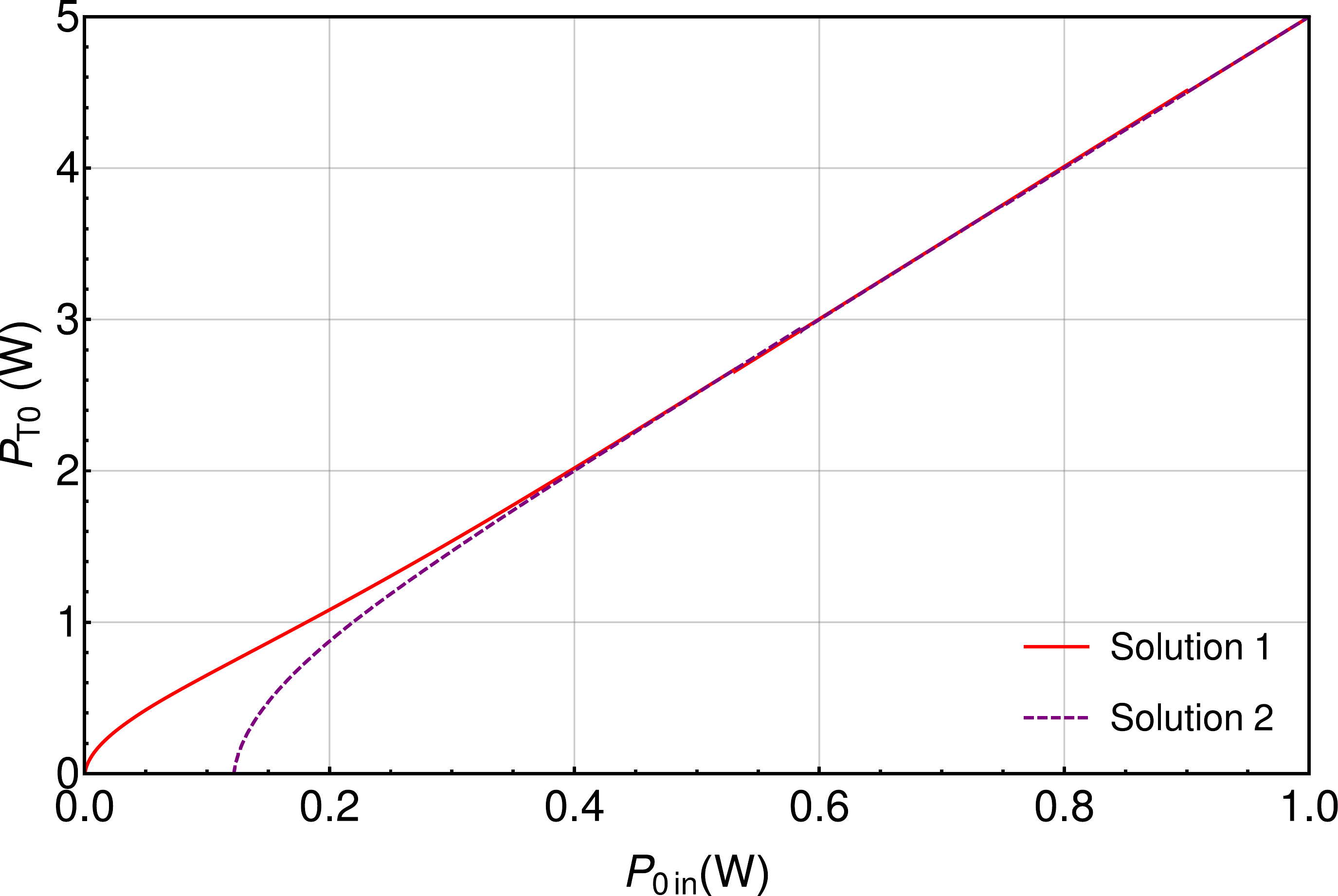}
\caption{\label{fig:TROPO_PT0} $P_{T0}$ as a function of $P_{0in}$. Solution 1 and Solution 2 are the results of $P_{T0}(P_{0in})$ when you carry the calculation with each of the two roots of $P_0(0)$.}
\end{figure}

From these two possible results, we can numerically find two solutions for $P_{T0}$ for a given $P_{0in}$
%Using these results in~Eq. (\ref{resultado_ganho_1}) and~ Eq. (\ref{eq:sqrtP0out}) we have two equations that relates $P_{0in}$ and $P_{T0}$ for each set of $P_0(0)$ and $P_1(0)$. 
%Let's call those equations $f_{mn}(P_{T0},P_{0in})=0$, we have $f_{11}(P_{T0},P_{0in})$ and $f_{12}(P_{T0},P_{0in})$ corresponding to the set $\left[ P_{0sol1}(0),P_{1sol1}(0)\right]$, $f_{21}(P_{T0},P_{0in})$ and $f_{22}(P_{T0},P_{0in})$ corresponding to the set $\left[ P_{0sol2}(0),P_{1sol2}(0)\right]$.
%Solve $f_{mn}(P_{T0},P_{0in})=0$ to find $P_{T0}$ as a function of $P_{0in}$ is a difficult task even with computational methods. However, $P_{0in}$ as a function of $P_{T0}$ have a analytic answer. By carrying the calculations we found out that $P_{0in}(P_{T0})$ have equivalent results for $f_{11}(P_{T0},P_{0in})$ and $f_{21}(P_{T0},P_{0in})$ (solution 1) and  also for $f_{12}(P_{T0},P_{0in})$ and $f_{22}(P_{T0},P_{0in})$ (solution 2).
As an example, Fig. \ref{fig:TROPO_PT0} shows those two solutions behave for $R_0=0.95$ and $R_1=0.80$. Since $P_{T0}>0$ for $P_{0in}>0$, the second solution can be discarded, and just the first one is used for the evaluation of the output fields, feeding back the values into Eqs. (\ref{resultado_ganho_1}) and (\ref{eq_P_1out}).

%To decide between solutions 1 and solution 2 we compared the two results with what is physically expected from the system. $P_{0in}$ as a function of $P_{T0}$ must cross the origin of the graphic and that just happens for solution 1.%This figure was made from a table which the first column are the $P_{0in}(P_{T0})$ and the second column $P_{T0}$. 

%If we adjust a curve to that plot, we have a curve that represents $P_{0}(P_{0in})$ in the interval of the adjustment. We made this procedure for any of the cavities we want to analyse. Then, we can replace the adjust of $P_{T0}$ on~\autoref{resultado_ganho_1} and then,~\autoref{resultado_ganho_1} on~ Eq. \ref{eq_P_1out}. First, in~\autoref{fig:R0difR1}(a) we present the model considering different possibilities for $R_{0}$ while keeping $R_{1}$ fixed. When we increase the value of $R_{0}$ both threshold and signal output power decrease. In~\autoref{fig:R0difR1}(b) we compare the model presented in this section with the model used in~\autoref{TROPO_R} for $R=75\%$,$R=85\%$ and $R=95\%$, showing an agreement between both models. With this treatment extension, we have presented a very versatility TROPO source using a $\chi^3$ gain medium.

%%%old form
%It is interesting to analyse the situation in which the cavity has different reflection coefficients for the resonant beams. Considering the TROPO with a $\chi^{(3)}$ gain medium we can extended the treatment in section~\ref{TROPO_R} using a cavity with reflection coefficients $R_0$ for the pump beam and $R_1$ for the signal and idler beams. 
%The self consistences equations for the pump beam are still the same as Eqs.~(\ref{eq:sqrtP00}) and~(\ref{eq:sqrtP0out}) but with $r$ and $t$ replaced by $r_0$ and $t_0$. 
%The dependence of $P_{1out}$ in terms of the main cavity parameters, $P_{0in}$, $R_0$ and $R_1$ as function of the the input pump power is showed in~Fig. \ref{fig:P1out_R0difR1}. 

\begin{figure}[ht]
	\centering
	\includegraphics[scale=0.3]{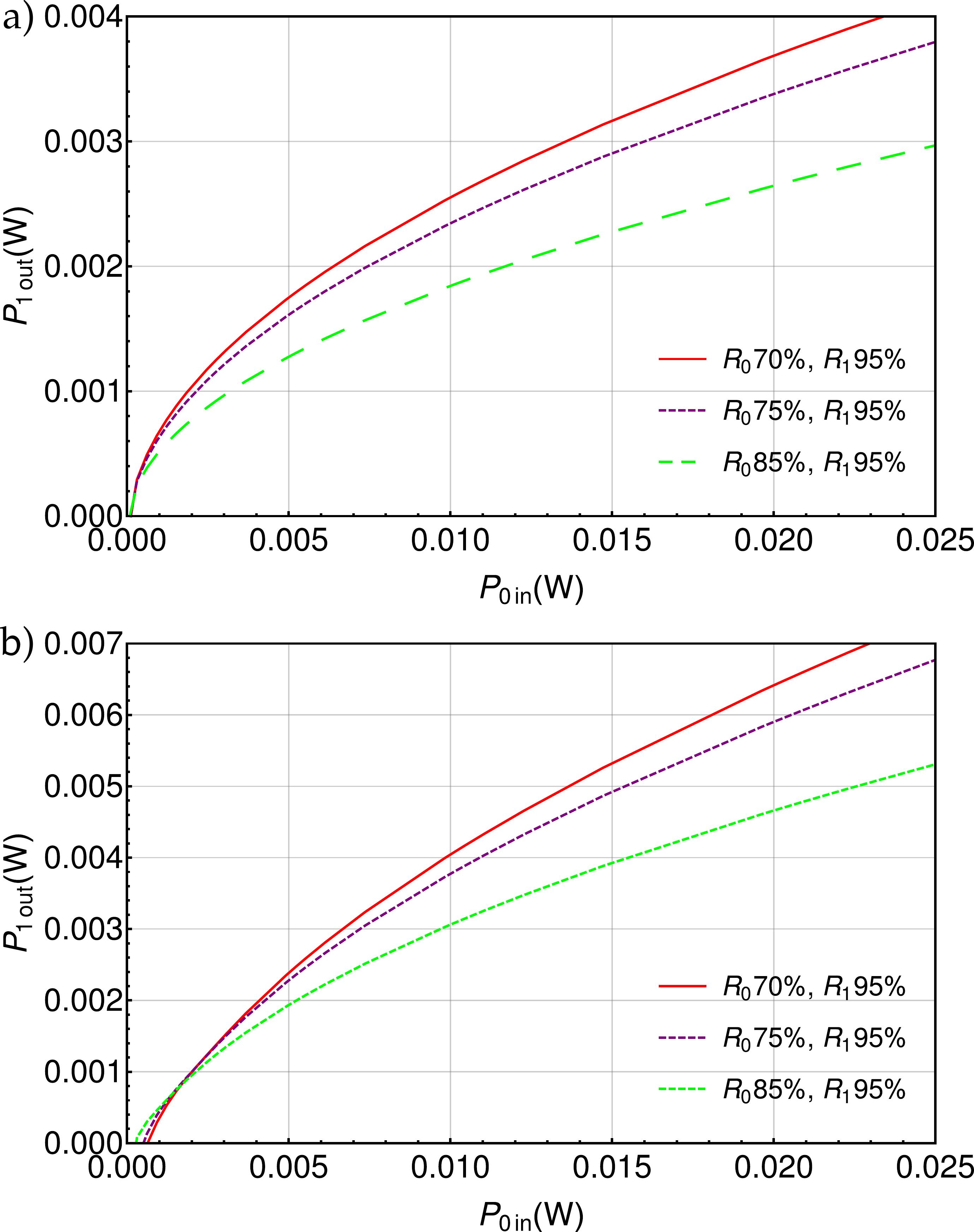}
 \caption{\label{fig:P1out_R0difR1} TROPO $R_1 \neq R_0$. $P_{1out}$ in function of $P_{0in}$ for different reflectivity coefficients $R_{0} = \{75\%, 85\%, 95\%\}$ and $R_{1} = 75\%$. a)  $\chi^{(2)}$ TROPO  b) $\chi^{(3)}$ TROPO.}
\end{figure}

We present the output power of the TROPO considering different possibilities for $R_{0}$ ($75\%$,$85\%$ and $95\%$) while keeping $R_{1}=95\%$ fixed. In Fig. \ref{fig:P1out_R0difR1}a we present the model used in~\cite{Debuisschert:93} for a $\chi^{(2)}$ TROPO, that it is a widely used description, for comparison with the results of Fig. \ref{fig:P1out_R0difR1}b, where we present the $\chi^{(3)}$ TROPO output power behavior. 
One can notice that the curve deviates from the parabolic response for $\chi^{(2)}$ media, but  as in Fig.~\ref{fig:TROPOMEDIO}, the qualitative behavior is similar.

\section{Input-output relations in the open cavity\label{transform}}

The explicit derivation of Eq. (\ref{cavityBS}), following the system described in Fig. \ref{cavfp}, follows the description presented in \cite{MUNOZ2018}. 
 The coupling mirror has reflection and transmission coefficients $r_{n}=\sqrt{R_n}$ and $t_{n}=sqrt{1-R_n}$ for each carrier, and we assume that one of the mirror as a reflection coefficient $r'_{n}$ and transmission coefficient $t'_{n}$ accounting for  spurious losses.
% These coefficients can be conveniently described by loss parameters $\gamma_{n}$ and $\gamma'_{n}$ as
%\begin{eqnarray}
%r_{n}= e^{-\gamma_{n}},
%\quad\quad
%t_{n}= (1-r'^{2}_{n})^{1/2},\nonumber\\
%r'_{n}= e^{-\gamma'_{n}},
%\quad\quad
%t'_{n}= (1-r'^{2}_{n})^{1/2}\label{m19a}.
%\end{eqnarray}
%The total loss in a round trip can be directly evaluated from $\gamma^t_{n}=\gamma_{n}+\gamma'_{n}$.
%The formalism adopted here  remains valid even in the open cavity regime, enabling the treatment in the limit where the cavity is completely open for one of the modes, as in the case of a doubly resonant OPO \cite{opencavity}.
The equations relating each field operator inside and outside the cavity are given by the beam splitter transformations
\begin{align}
\vec{\mathbf{A}}_{\textsc{R}}&= \textbf{R}\vec{\mathbf{A}}_{\text{in}}+\textbf{T}\vec{\mathbf{B}}', &
\vec{\mathbf{B}}&= \textbf{T}\vec{\mathbf{A}}_{\textrm{in}}- \textbf{R}\vec{\mathbf{B}}',&\label{m13}\\[2mm]
\vec{\mathbf{A}}_{\textrm{T}}&= \textbf{R}'\vec{\mathbf{A}}_{\nu}+\textbf{T}' \vec{\mathbf{C}},&
\vec{\mathbf{C}}'&= \textbf{T}'\vec{\mathbf{A}}_{\nu}- \textbf{R}'\vec{\mathbf{C}},&\label{m14}
\end{align}
with
\begin{align}
\label{rt}
\textbf{R}&=\text{diag}\big(r_{0}~r_{0}~r_{1}~r_{1}~r_{2}~r_{2}~ r_{0}~r_{0}\cdots\big),\notag\\[2mm]
\textbf{T}&=\text{diag}\big(t_{0}~t_{0}~t_{1}~t_{1}~t_{2}~t_{2}~ t_{0}~t_{0}\cdots\big),\notag\\[2mm]
\textbf{R}'&=\text{diag}\big(r'_{0}~r'_{0}~r'_{1}~r'_{1}~r'_{2}~r'_{2}~ r'_{0}~r'_{0}\cdots\big),  \notag\\[2mm]
\textbf{T}'&=\text{diag}\big(t'_{0}~t'_{0}~t'_{1}~t'_{1}~t'_{2}~t'_{2}~ t'_{0}~t'_{0}\cdots\big),
\end{align}
with the vector fields changed from the symmetric/antisymmetric basis into the basis of the sideband operators $\vec{\mathbf{A}}=(\hat{a}^{(0)}_{\omega_{0}+\Omega} ~\hat{a}^{(0)\dagger}_{\omega_{0}+\Omega}~\cdots ~\hat{a}^{(0)}_{\omega_{0}-\Omega} ~\hat{a}^{(0)\dagger}_{\omega_{0}-\Omega}~\cdots )^{T}$.

The round trip of the fields inside the cavity will account for both parametric gain $\mathbf{G'}$ (Eq. \ref{gain}, properly transformed into the sideband mode basis) and additional phase, leading to the transformation
\begin{eqnarray}
\vec{\mathbf{C}}= e^{-i \boldsymbol{\varphi}}\mathbf{G'} \vec{\mathbf{B}}, \quad \quad
\vec{\mathbf{B}}'= e^{-i\boldsymbol{\varphi}}\mathbf{G'}  \vec{\mathbf{C}}'.\label{m19}
\end{eqnarray}
The phase vector
\begin{equation}
\boldsymbol{\varphi}=\boldsymbol{\varphi}(\Omega) \oplus \boldsymbol{\varphi}(-\Omega),\label{m20}
\end{equation}
with,
\begin{eqnarray*}
\boldsymbol{\varphi}(\Omega)=  \text{diag}\big(
\varphi_{\Omega}^{(0)},-\varphi_{\Omega}^{(0)}
\varphi_{\Omega}^{(1)}, -\varphi_{\Omega}^{(1)},
\varphi_{\Omega}^{(2)}-\varphi_{\Omega}^{(2)} \big),
\end{eqnarray*}
gives a different contribution for each sideband depending of the frequency shift $\Omega$% and on the carrier frequency $\omega_n$
\begin{eqnarray}\label{phase}
\boldsymbol{\varphi}_{\Omega}^{(n)}=
\dfrac{\Omega}{2\,\text{FSR}_{n}}.
\end{eqnarray}
where we consider exact resonance of the carrier mode and $\text{FSR}_{n}=c/2L$ as the free spectral range for the mode $n$.
%Evidently the effective phase contribution will depend on the detuning between the carrier and the nearest cavity mode $\omega^c_n$, an integer multiple of $2\pi \text{FSR}_{n}$, given by  $\Delta_n=\omega_n -\omega^c_n$.
%An important point related to the evolution of the sidebands should be noticed. Each operator will undergo a different phase evolution, depending on their frequency. That will mix symmetric and antisymmetric modes, even for null carrier detuning, since upper and lower sidebands will, in this case, undergo opposite phase evolutions. This is the cause of the correlations between symmetric and antisymmetric modes observed in \cite{hexaopo}.

Combining beam splitter transformation, phase evolution and gain, expressed in Eqs.(\ref{m13}--\ref{m19})  we obtain the linear transformation, Eq. \ref{cavityBS}
with the coupling matrices given by
\begin{eqnarray}
\textbf{R}_{\chi}&=&\textbf{R}-\textbf{T}e^{-i\boldsymbol{\varphi}}
\mathbf{G}(\chi)\textbf{R}'e^{-i\boldsymbol{\varphi}}\mathbf{G}(\chi)
\mathbf{D}(\chi)\textbf{T},\label{Eq:26}\\[2mm]
\textbf{T}'_{\chi}&=& \textbf{T}e^{-i\boldsymbol{\varphi}}\mathbf{G}(\chi) \left[\textbf{I} +\textbf{R}'
e^{-i\boldsymbol{\varphi}}\mathbf{G}(\chi)\mathbf{D}(\chi)
\textbf{R}e^{-i\boldsymbol{\varphi}}\mathbf{G}(\chi) \right] \textbf{T}',\nonumber\\\label{Eq:27}
\end{eqnarray}
and
\begin{eqnarray}
\mathbf{D}(\chi)=\Big( \textbf{1}-\textbf{R}e^{-i\boldsymbol{\varphi}}\mathbf{G}(\chi) \textbf{R}'e^{-i\boldsymbol{\varphi}}\mathbf{G}(\chi)\Big)^{-1},\label{Eq:28}
\end{eqnarray}
with special care in the basis transformation, from the sideband description (useful for phase propagation given by Eq. \ref{phase}) to the symmetric/antisymmetric combination (useful for parametric gain given by Eq. \ref{gain}), as done in \cite{MUNOZ2018}.

\section{Correlations between symmetric and antisymmetric basis\label{App_CV_sym_antsym}}

\begin{figure}[!ht]
\centering
\includegraphics[scale=0.25]{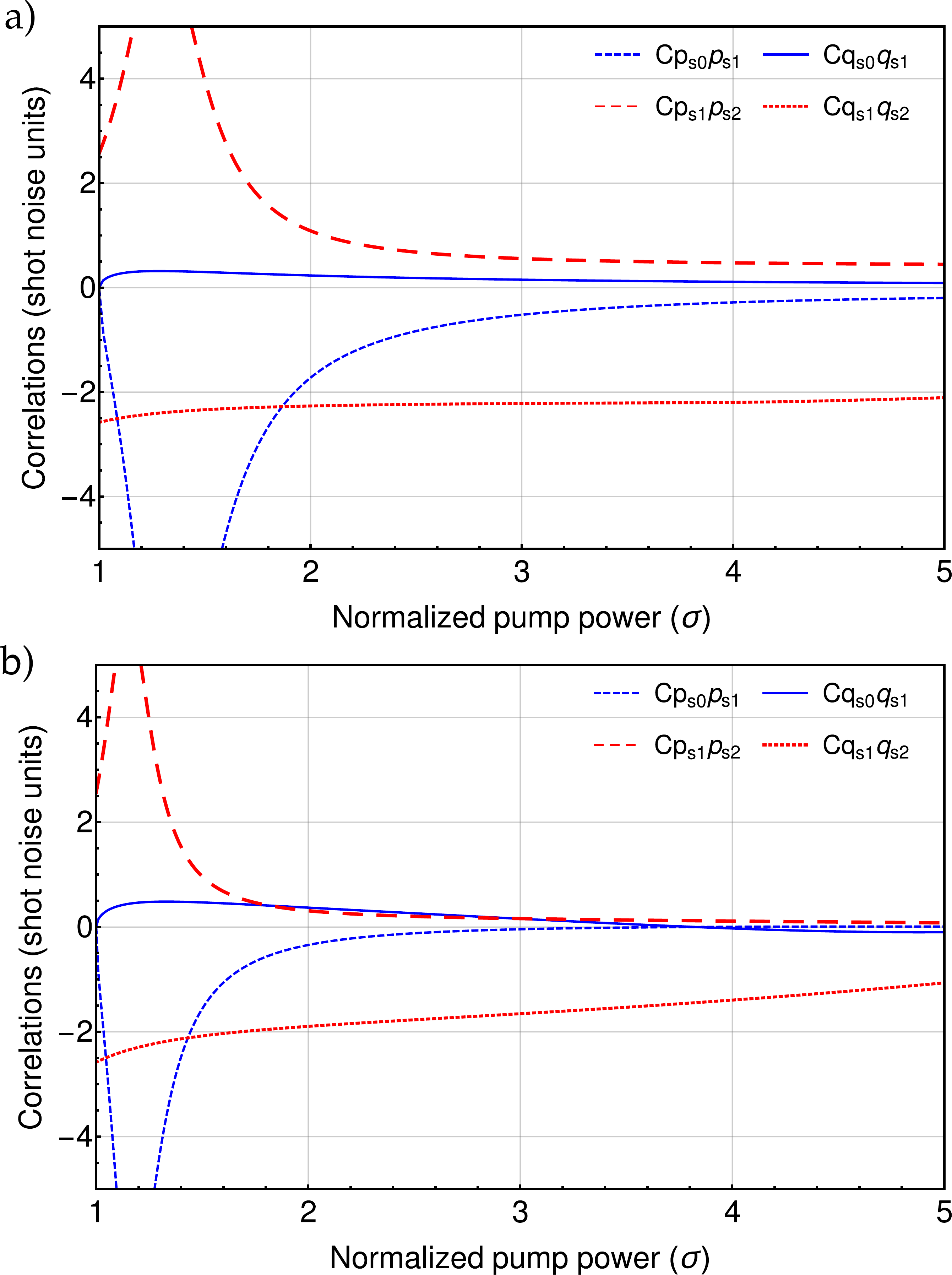}
\caption{\label{CorrSimchi23} DROPO: Non-diagonal elements of $\textbf{V}_{s}$ as a function of  $\sigma$ for $R_{1}=85\%$ and $\Omega = 0.5 BW$. Considering in a) a $\chi^{(2)}$ gain medium and in b) a $\chi^{(3)}$ gain medium.}
%\end{figure}
%\begin{figure}[h!]
%\centering
\includegraphics[scale=0.25]{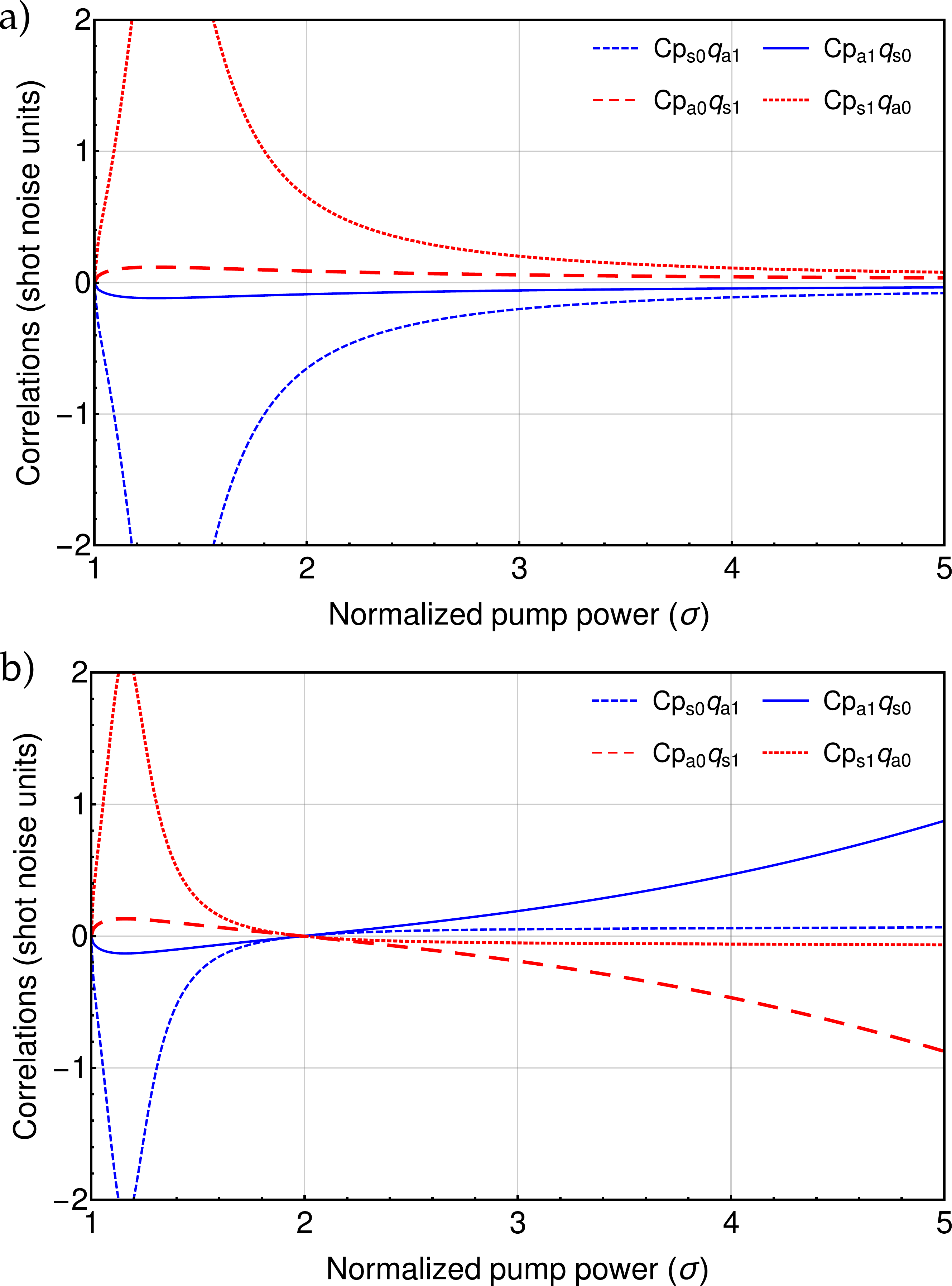}
\caption{\label{chi2chi3_correlation_simantisim} DROPO: Correlations between the symmetric and antisymmetric basis  as a function of $\sigma$ for $R_{1}=85\%$ and $\Omega = 0.5 BW$. Considering in a) a $\chi^{(2)}$ gain medium and in b) a $\chi^{(3)}$ gain medium.}
\end{figure}

\begin{figure}[!ht]
\centering
\includegraphics[scale=0.25]{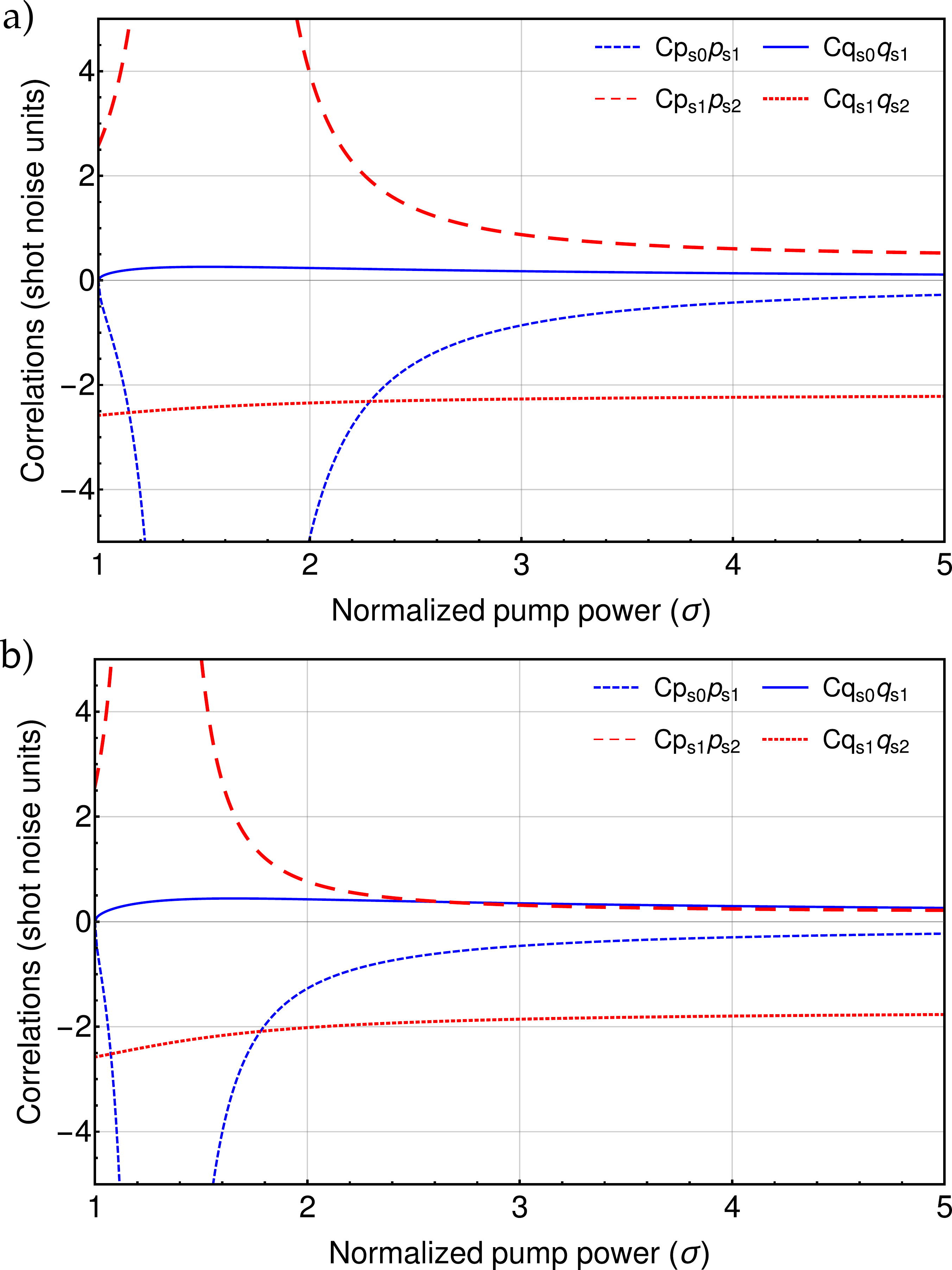}
\caption{\label{fig:TROPOCorrelacaoSimetrica} TROPO: Non-diagonal elements of $\textbf{V}_{s}$  as a function of $\sigma$ for $R_{1}=85\%$ and $\Omega = 0.5 BW$. Considering in a) a $\chi^{(2)}$ gain medium and in b) a $\chi^{(3)}$ gain medium.}
%\end{figure}
%\begin{figure}[h!]
%\begin{center}
\includegraphics[scale=0.25]{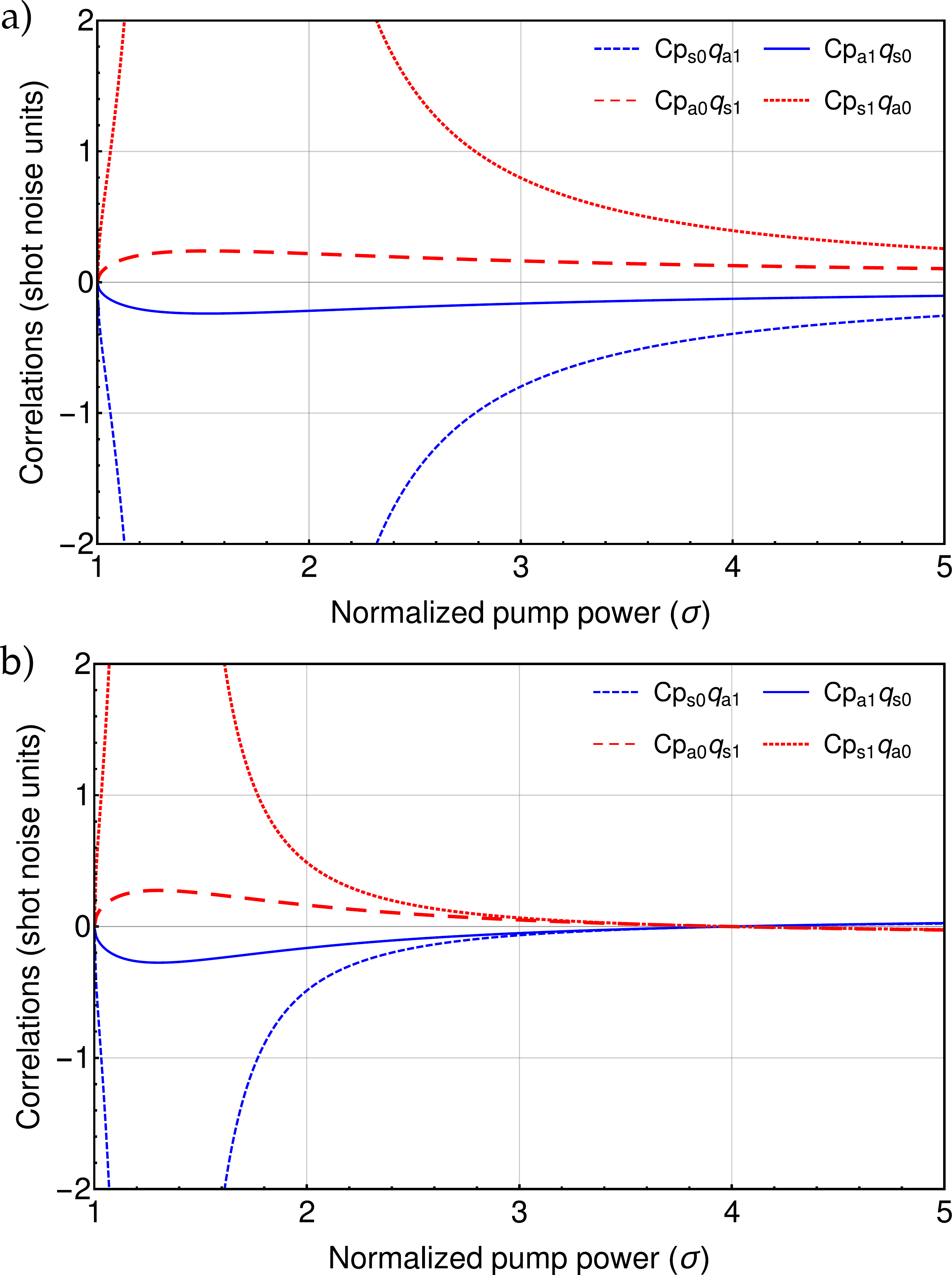}
\caption{TROPO: Correlations between the symmetric and antisymmetric basis  as a function of  $\sigma$ for $R_{1}=85\%$ and $\Omega = 0.5 BW$. Considering in a) a $\chi^{(2)}$ gain medium and in b) a $\chi^{(3)}$ gain medium.
\label{fig:TROPOCorrSimAnt}}
%\end{center}
\end{figure}

The non-diagonal terms of the covariance matrix~ Eq. (\ref{eq:matCovar_main})  in the symmetric basis are presented in Fig. \ref{CorrSimchi23}. The terms in antisymmetric basis are omitted since they are equal to the terms in symmetric basis under a rotation of $\pi/2$ in one of the field modes \cite{PhysRevA.88.052113}. The twin beams present correlation between the amplitude quadratures, $C p_{s1} p_{s2}$ and anti-correlation between the phase quadratures, $C q_{s1} q_{s2}$, for all values of $\sigma$ in both cases, as can be seen in  Fig. \ref{CorrSimchi23}a for a $\chi^{(2)}$ gain medium and Fig. \ref{CorrSimchi23}b for a $\chi^{(3)}$ gain medium. 
Both curves present a peak in the amplitude correlations ($Cp_{s0}p_{s1}$ and $Cp_{s1}p_{s2}$) associated to the peak in the amplitude noise observed in Fig. \ref{VarSimchi23}. The general behavior is pretty similar for the TROPO case, as showed in Fig. \ref{fig:TROPOCorrelacaoSimetrica}.
The greatest difference among the distinct configurations is for the $\chi^{(3)}$ DROPO, where we can observe a flip on the pump-signal correlations for amplitude ($Cp_{s0}p_{s1}$) and phase quadratures ($Cq_{s0}q_{s1}$) at $\sigma\approx3.6$, associated with the apparent disentanglement between pump and signal observed in Fig. \ref{pptchi23}.

The cross-correlations terms between symmetric and antisymmetric field modes are showed in~Fig. \ref{chi2chi3_correlation_simantisim} for the DROPO and Fig. \ref{fig:TROPOCorrSimAnt} for the TROPO. The relevant term is $Cp_{s0}a_{s1}=-Cp_{s1}q_{a0}$ in most of the situations. Remembering that the phases of the antisymmetric basis are rotated \cite{PhysRevA.88.052113}, it is this the leading term associated to the apparent loss of entanglement in the tripartite case, that is recovered once the full covariance matrix is taken into account. A curious feature appears only in the $\chi^{(3)}$ case: at $\sigma=2$ (for DROPO) or $\sigma=4$ (for TROPO) the correlation goes to zero. That leads to a perfect decoupling of the symmetric and antisymmetric modes, as observed by the unitary value of the symplectic eigenvalue of the partially transposed matrix observed in Figs. \ref{fig:DROPOPPThexa} and \ref{fig:TROPOPPThexa}.

%\newpage

\end{document}